\documentclass[twocolumn]{aastex631}

\graphicspath{{./}{figures/}}
\usepackage{amsmath}
\usepackage{threeparttable}
\usepackage{graphicx}        
\usepackage{CJK}
\usepackage{tablefootnote}

\newcommand\sref[1]{\hyperref[#1]{\S~\ref*{#1}}}
\newcommand\fref[1]{\hyperref[#1]{Fig.~\ref*{#1}}}
\newcommand\Eqref[1]{Eq.~(\hyperref[#1]{\ref*{#1}})}
\newcommand\eeqref[1]{Eq.~\hyperref[#1]{\ref*{#1}}}
\newcommand\tref[1]{\hyperref[#1]{Table~\ref*{#1}}}
\newcommand\aref[1]{\hyperref[#1]{Appendix~\ref*{#1}}}

\shorttitle{To Survive or to Shatter}
\shortauthors{Roy et al.}

\begin{document}
\begin{CJK}{UTF8}{mj}
\title{{To Survive or to Shatter: The Impact of Cosmic Rays on the Fate of Stripped Cold Clouds}}

\correspondingauthor{Manami Roy}
\email{roy.516@osu.edu}
\author{Manami Roy}
\affiliation{Center for Cosmology and Astro Particle Physics (CCAPP), The Ohio State University, 191 W. Woodruff Avenue, Columbus, OH 43210, USA}
\affiliation{Department of Astronomy, The Ohio State University, 140 W. 18th Ave., Columbus, OH 43210, USA}
\author{Kung-Yi Su}
\affiliation{Black Hole Initiative, Harvard University, 20 Garden St., Cambridge, MA 02138, USA}
\affiliation{Department of Physics and Astronomy and CIERA, Northwestern University, 2145 Sheridan Road, Evanston, IL 60208, USA}
\author{Stephanie Tonnesen}
\affiliation{Center for Computational Astrophysics, Flatiron Institute, 162 5th Ave, New York, NY 10010, USA}
\author{Yue Samuel Lu}
\affiliation{Department of Astronomy and Astrophysics, University of California, San Diego, La Jolla, CA 92093, USA}
\affiliation{Physics Department, University of California, San Diego, La Jolla, CA 92093, USA}
\author{ Cameron Hummels}
\affiliation{TAPIR, California Institute of Technology, Mailcode 350-17, Pasadena, CA 91125, USA}
\author{Sam B. Ponnada}
\affiliation{TAPIR, California Institute of Technology, Mailcode 350-17, Pasadena, CA 91125, USA}

\begin{abstract} \label{abstract}
Does cosmic ray (CR) pressure matter for the circumgalactic medium (CGM)? Despite growing interest, this remains a debated question, complicated by limited observational constraints and differing implementations of CR physics in simulations. While prior studies suggest that CRs influence the thermal and dynamical state of the CGM, their role in shaping cold gas structures remains underexplored. This paper investigates how CRs affect ram-pressure stripped cold gas clouds originating from satellite galaxies in a Milky Way-like halo. Using high-resolution simulations with varying CR energy densities, we find that CRs can significantly modify the size and survival of stripped clouds. Specifically, CR pressure puffs up the cold clouds, increasing their surface area and enabling more efficient mixing-layer cooling, allowing them to grow in mass. This enhanced growth results in higher cold gas inflow rates into the central galaxy, leading to an increase in the star formation rate compared to the no-CR case at a later time. Moreover, CRs can boost the total cold gas mass in the CGM by up to a factor of four. These effects are most pronounced in simulations where the CR energy density is in equipartition with the thermal gas. Our results demonstrate that CRs can play a critical role in regulating the cold phase of the CGM contributed by satellites and therefore their ability to feed galaxies.    
\end{abstract}

\section{Introduction}

The diffuse gaseous halo surrounding a galaxy, known as the circumgalactic medium (CGM), acts as a crucial interface between the galactic disk and the intergalactic medium \citep[See review by][]{Tumlinson2017, F&Oh2023}. The CGM is composed of multiphase gas, and is a substantial reservoir for cold gas (T $< 10^4$ K), which composes up to $\sim30-50$\% of a galaxy’s total baryonic matter \citep{Werk2014}. The cold phase of the CGM is thought to be an important source of star-formation fuel, as it can accrete onto the galactic disk and sustain ongoing star formation. Understanding the origin, survival, and transport of this cold gas is therefore crucial for constructing a comprehensive picture of galaxy evolution. However, prime questions about this cold phase still persist in the field: \textit{How does this gas accumulate in the outer regions of the CGM \citep[e.g.][]{Tumlinson2013,Keeney2018}? Why do quenched galaxies, which no longer form stars, still harbor large reservoirs of cold gas \citep[e.g.][]{Berg2019}? Why do COS-Halos observations suggest that the density of this cold gas is an order of magnitude lower than expected if it were in thermal pressure equilibrium with the surrounding hot gas \citep{Werk2014}?}

These questions have sparked significant interest and speculation in recent years. Two primary scenarios have been proposed to explain the origin of cold gas in the CGM: either it originates ex-situ, e.g., in galactic disks, satellite galaxies, or the intergalactic medium \citep[e.g.][]{Li2020, Fielding2020, Gronke2018, Roy2024} and has been transported into the halo, or it forms in-situ, through cooling and condensation processes within the CGM itself \citep[e.g.][]{McCourt2011, Sharma2012, Voit2015, Roy2021}. 
Satellite galaxies can contribute substantial amounts of cold gas to the CGM through both ram pressure stripping and induced cooling in the mixing layer of the stripped gas, with both mechanisms playing comparable roles \citep[e.g.][]{Roy2024, Damle2025}. 
Cold gas in the CGM can also form in situ via thermal instability, wherein small perturbations in hot, gravitationally stratified gas can trigger runaway cooling and condensation. This process, initially proposed in the context of galaxy clusters, is now recognized as an important mechanism for generating cold CGM gas \citep[e.g.][]{McCourt2011, Sharma2012, Voit2015} and regulating the baryon cycle. 
Therefore, both in-situ formation and external transport contribute to the presence of cold gas in the CGM, though their relative importance varies across galaxies. While cold gas is expected to be short-lived in the hot halo environment  \citep[e.g.][]{Mckee1975, Zhang2017}, observations of multiphase outflows \citep[e.g.][]{Flue2021, Ashley2020} and recent theoretical works \citep[e.g.][]{Li2020, Fielding2020, Gronke2018} suggest that cold gas can survive and even grow in hot winds. However, it remains uncertain whether these mechanisms can fully account for the observed cold gas mass in the CGM.

While multiple pathways exist for the origin of the cold phase in the CGM, the unexpectedly low densities of this gas inferred from UV absorption-line observations \citep{Werk2014} remain a topic of active debate. However, it hints towards non-thermal pressure support, with cosmic rays (CRs) emerging as a prime suspect. However, the influence of CRs on cold gas has been investigated in only a few studies. \cite{Sharma2010} suggested that adiabatic CRs can suppress cold gas formation by reducing the growth of thermal instabilities. A previous study by \cite{Butsky2020} has shown that CRs can influence the thermal instability-driven formation of cold gas by providing non-thermal pressure support. This additional pressure reduces the compressibility of the gas, enabling it to cool at lower densities and leading to the formation of larger cold structures. Their simulations that include CR physics often produce extended cold gas clouds or mist-like configurations, in contrast to the small, dense cloudlets typically formed in their absence. In some cases, they find that CR pressure can increase the cold gas mass in the halo by inhibiting the inward precipitation of condensed gas toward the galactic center. However, the impact of CRs on this ex-situ cold phase from satellite or IGM remains largely unexplored. 

In this paper, we investigate the impact of CRs on ram-pressure stripping and explore these key questions: Do CRs affect the stripping rate from the satellites? What role do CRs play in the evolution of the mass and size of stripped cold clouds?  Do CRs impact the survival of cold clouds and their subsequent infall and feeding of the central galaxy? In Section \ref{S:methods}, we discuss the methodology of our simulation, where we describe the initial conditions of our simulation setup (Section \ref{S:ic}) along with our definition of cold CGM gas for our analysis (Section \ref{S:Def}). In Section \ref{S:Result}, we demonstrate our results from our analysis of the simulations, where we discuss the size and mass distribution of ram-pressure stripped clouds (Section \ref{S:size}), mass (Section \ref{S:mass}), covering fraction (Section \ref{S:cv}) of the cold gas in the CGM and, their inflow rate (Section \ref{S:inflow}). Finally, we discuss in more detail the impact of CR physics in Section \ref{S:Discussion}. To conclude, we summarize our results and discuss future work in Section \ref{S:Conclusion}. 


\section{Methodology}
\label{S:methods}
We perform our simulations using the {\textsc{gizmo}} code\footnote{Publicly available at \href{http://www.tapir.caltech.edu/~phopkins/Site/GIZMO.html}{\textsc{gizmo}}} \citep{2015MNRAS.450...53H}, operating in its meshless finite mass (MFM) mode. This Lagrangian, mesh-free Godunov method combines the strengths of both smoothed-particle hydrodynamics (SPH) and grid-based techniques. The numerical framework and validation tests for various physics modules -- including hydrodynamics, self-gravity \citep{2015MNRAS.450...53H}, magnetohydrodynamics (MHD) \citep{2016MNRAS.455...51H, 2015arXiv150907877H}, anisotropic conduction and viscosity \citep{2017MNRAS.466.3387H, 2017MNRAS.471..144S}, and cosmic rays \citep{chan:2018.cosmicray.fire.gammaray} -- are detailed in a series of associated papers.

All simulations implement the FIRE-2 (Feedback in Realistic Environments) model, which incorporates a comprehensive treatment of the multiphase interstellar medium (ISM), star formation, and stellar feedback \citep{hopkins:sne.methods, 2017arXiv170206148H}. Gas cooling is modeled over a wide temperature range ($10$-$10^{10}$ K), accounting for processes such as photoelectric and photoionization heating, collisional and Compton cooling, atomic and molecular transitions, fine-structure emission, and recombination.

Star formation is modeled using sink particles that form only in dense, molecular gas regions exceeding $100\,{\rm cm^{-3}}$, where local self-gravity and self-shielding dominate. Newly formed star particles inherit the metallicity of their progenitor gas and represent single-age stellar populations. Stellar feedback is implemented based on IMF-averaged quantities derived from {\small STARBURST99} \citep{1999ApJS..123....3L}, assuming a Kroupa initial mass function (IMF) \citep{2002Sci...295...82K}.

The feedback model includes multiple physical channels: Radiative feedback, including photoionization, photoelectric heating, and radiation pressure (single and multiple scattering), tracked across five spectral bands (ionizing, FUV, NUV, optical-NIR, and IR); Stellar winds, accounting for mass, energy, momentum, and metal injection from OB and AGB stars; Supernovae feedback, from both Type II and Ia SNe, including prompt and delayed explosions, with the corresponding thermal and kinetic energy, mass, and metal injection into the surrounding ISM. All simulations incorporate MHD, fully anisotropic thermal conduction, and viscosity using Spitzer-Braginskii coefficients.

For simulations with cosmic rays (CRs), we include anisotropic streaming and diffusion along magnetic field lines. CR streaming occurs at the maximum of the local Alfven speed or sound speed and includes a thermalization loss term \citep{Uhlig2012}. CR diffusion is modeled using a fixed diffusivity of $\sim10^{29}\,\rm cm^2\,s^{-1}$, for our fiducial run, coupled with adiabatic exchanges with the thermal gas and losses via Coulomb and hadronic interactions \citep{Guo2008}. We model a single GeV energy bin, appropriate for the dominant CR pressure component, and treat CRs as ultra-relativistic particles. CRs are injected by SNe, with 10\% of the explosion energy allocated to CRs, consistent with previous studies \citep{Pfrommer2017a, Pfrommer2017b}. Additional implementation details and physical modeling of CRs are described in \cite{Su2019, chan:2018.cosmicray.fire.gammaray}.

\subsection{Initial Conditions}
\label{S:ic}
Our initial setup follows the methodology described in \citet{Su2019}, \citet{Roy2024}, and \cite{Roy2024b}. To ensure a stable and quasi-equilibrium CGM environment, the simulation domain is extended to three times the virial radius of the host halo. The system is then evolved adiabatically (i.e., without radiative cooling or star formation) for 4.5 Gyr to allow any initial transients to relax before introducing satellite galaxies. Simulation parameters and detailed properties are provided in Table \ref{t:runs}. In this study, we focus exclusively on the \texttt{m12} host halo, representing a Milky Way-mass galaxy with a total mass of approximately $1.8\times10^{12}\, {\rm M_{\odot}}$ with either 2 or 20 satellite galaxies, each embedded in a dark matter halo of $2\times10^{10} \,{\rm M_{\odot}}$ (denoted as {\texttt m10}) and $2\times10^{9} \, {\rm M_{\odot}}$ (denoted as {\texttt m09}), respectively. Additionally, the m09 satellite has a gas mass of $7 \times 10^7\,M_\odot$ and a gas disk radius of 0.87\,kpc, while the m10 satellite has a gas mass of $4.2 \times 10^8\,M_\odot$ and a gas disk radius of 2.1\,kpc.

We initialize the dark matter (DM) halo, stellar bulge, central black hole, and gas+stellar disk of the {\texttt m12} galaxy following the methodology outlined in \citet{1999MNRAS.307..162S,2000MNRAS.312..859S}. The initial gas metallicity decreases radially from solar ($Z = 0.02$) to $Z = 0.001$, following the profile:
\begin{equation}
    Z(r) = 0.02\,\left(0.05 + \frac{0.95}{1 + (r / 20\,{\rm kpc})^{1.5}}\right).
\end{equation}
Magnetic fields are initialized in the azimuthal direction with a radial dependence:
\begin{equation}
    |{\bf B}(r)| = \frac{0.03\,\mu{\rm G}}{1 + (r / 20\,{\rm kpc})^{0.375}}.
\end{equation}
Further details of the initial setup can be found in the Methodology section of \citet{Roy2024}.

For simulations that include CRs, we adopt three different initialization schemes. In the high-CR runs 
(\texttt{CR\_h}), CR energy density are initialized in equipartition with the thermal energy density. To preserve the total pressure and maintain hydrostatic equilibrium, we reduce the initial gas temperature by 2/3 and allocate the remaining energy to CRs. In the mid-CR runs (\texttt{CR\_m1,2}), we consider that CRs are initially 1/10 and 1/4 th of the thermal energy. In the low-CR runs (\texttt{CR\_l}), CRs are initialized such that the CR energy density is in equipartition with the magnetic energy density, which is approximately five orders of magnitude lower than the thermal energy density. In Table \ref{t:runs}, we list all of our runs with a full description of their initial pressure and CR transport parameters like diffusion coefficient and magnetic field.    

\subsection{Definition and Classification of Cold CGM Gas}
\label{S:Def}
In this section, we describe all of our definitions used to classify cold gas throughout the simulation.  We note that they are the same as in \citet{Roy2024} and have been tested extensively for their robustness. 
\begin{itemize}
\item \textbf{Cold Gas:} To identify the cold phase of the CGM in our simulation, we adopt a temperature threshold of $T < 3 \times 10^4$\,K. 
\item \textbf{CGM and ISM boundary:} To focus exclusively on the CGM and exclude the interstellar medium (ISM) of the central galaxy, we remove from our analysis all gas within 40\,kpc (approximately $0.15\,R_{\rm vir}$) of the host galaxy’s center.
\item \textbf{Satellite boundary:} To define the spatial extent of a satellite's influence, we consider the decline in its gas surface density, given by $\Sigma = \Sigma_0\, e^{-r/r_d}$, where $r_d$ is the stellar scale radius. The gas scale radius is defined as $r_{\rm gd} \sim 2.7\, r_d$, beyond which $\Sigma$ drops significantly \citep{Krav2013}. We define a radius of $6\,r_{\rm gd}$ from the satellite center (identified with the coordinates of its central black hole) as the boundary beyond which the satellite’s gravitational influence on the surrounding gas becomes negligible.

\end{itemize}

We further distinguish between different origins of cold gas using the Lagrangian MFM method implemented in the \textsc{GIZMO} code, which allows us to track individual gas particles across the simulation. This enables us to classify cold gas based on both its thermal evolution and its dynamical interaction with satellite galaxies. Using this framework, we classify cold gas in the CGM into four categories:

\begin{enumerate}
    \item \textbf{Cold gas inside the satellite:} Gas particles that originate within the satellite and remain within $6\,r_{\rm gd}$.
    
    \item \textbf{Stripped cold gas:} Gas particles that originate within the satellite but move beyond $6\,r_{\rm gd}$ while remaining below $3 \times 10^4$\,K. These particles are considered to be cold gas stripped from the satellite.
    
    \item \textbf{Host Gas cooled inside the satellite:} Gas particles initially associated with the host CGM that cool to $T < 3 \times 10^4$\,K and move to within the satellite’s effective radius ($r < 6\,r_{\rm gd}$). This refers to hot gas from the host halo that enters the satellite's gravitational potential well and subsequently cools within it.
    
    \item \textbf{Host Gas cooled outside the satellite:} Gas particles that begin as host CGM and cool below $3 \times 10^4$\,K without ever entering the satellite's defined boundary. This is the host halo gas that undergoes cooling outside the satellite, either within mixing layers of stripped cold gas or due to turbulence-induced cooling.
\end{enumerate}

In summary, we define the cold CGM using a combination of temperature thresholds, spatial cuts, and particle tracking. This allows us to isolate gas cooled by different origins, setting the stage for the quantitative analysis in the next section.

\begin{table*}
    \caption{Summary of all the runs: m09: 20 of m09 satellites and m10: 2 of m10 satellites; D29: D=$10^{29}$ cm$^{2}$/s and D28: D=$10^{28}$ cm$^{2}$/s} \label{t:runs}
    \begin{tabular}{||c|c|c|c|c|c|c||}
    \hline
    Set & Symbol  & Resolution  & P$_{\rm th}$ & P$_{\rm CR}$ & D & $\beta$ \\ [0.3ex] 
    \hline\hline
    \textbf{1. Fiducial m09 runs} & m09\_noCR & low & 1 & 0 & No & profile \\
    &m09\_noCR\_hr & high & 1 & 0 & No & profile  \\ 
     &m09\_CR\_l\_hr & high & 1 & profile & D29 & profile  \\ 
    &m09\_CR\_l & low & 1 & profile & D29 & profile  \\ 
    &m09\_CR\_m1 & low & 9/10 & 1/10 & D29 & profile \\ 
    &m09\_CR\_m2 & low & 3/4 & 1/4 & D29 & profile \\ 
    &m09\_CR\_h & low & 2/3 & 1/3 &  D29 & profile \\  
    \hline \hline 
    \textbf{2. Fiducial m10 runs} & m10\_noCR & low & 1 & 0 & No & profile \\
    & m10\_noCR\_hr & high & 1 & profile & No & profile \\
    & m10\_CR\_l & low & 1 & profile & D29& profile \\ 
    & m10\_CR\_m1 & low & 9/10 & 1/10  & D29& profile  \\
    & m10\_CR\_m2 & low & 3/4 & 1/4  & D29& profile  \\
    & m10\_CR\_h & low & 2/3 & 1/3 & D29 & profile \\ 
    \hline \hline
    \textbf{3. Other m09 runs} & m09\_CR\_h\_D28 & low & 2/3 & 1/3 & D28 & profile\\ 
    &m09\_CR\_h\_beta30 & low & 2/3 & 1/3 & D29 & 30 \\
    &m09\_CR\_m2\_overP & low & 1 & 1/4 & D29 & profile\\ 
    &m09\_noCR\_underP & low & 3/4 & 0 & No & profile \\ 
    &m09\_CR\_m2\_underP & low & 1/2 & 1/4 & D29 & profile \\
    \hline
    \end{tabular}
    \label{t:runs}
\end{table*}
\section{Results}
\label{S:Result}
The primary objective of this paper is to investigate the impact of CRs on the cold gas in the CGM that has been stripped from satellite galaxies by ram pressure in a Milky Way-like host halo. We begin by presenting visual insights from simulation snapshots to build physical intuition, followed by a quantitative analysis of the cold gas mass, size, and evolution under the influence of CRs. We restrict our analysis to up to and including the snapshot at 2.0 Gyr, as after that, the halo gas faces excessive cooling in the high-CR runs.  
\begin{figure*}
\includegraphics[width=\linewidth]{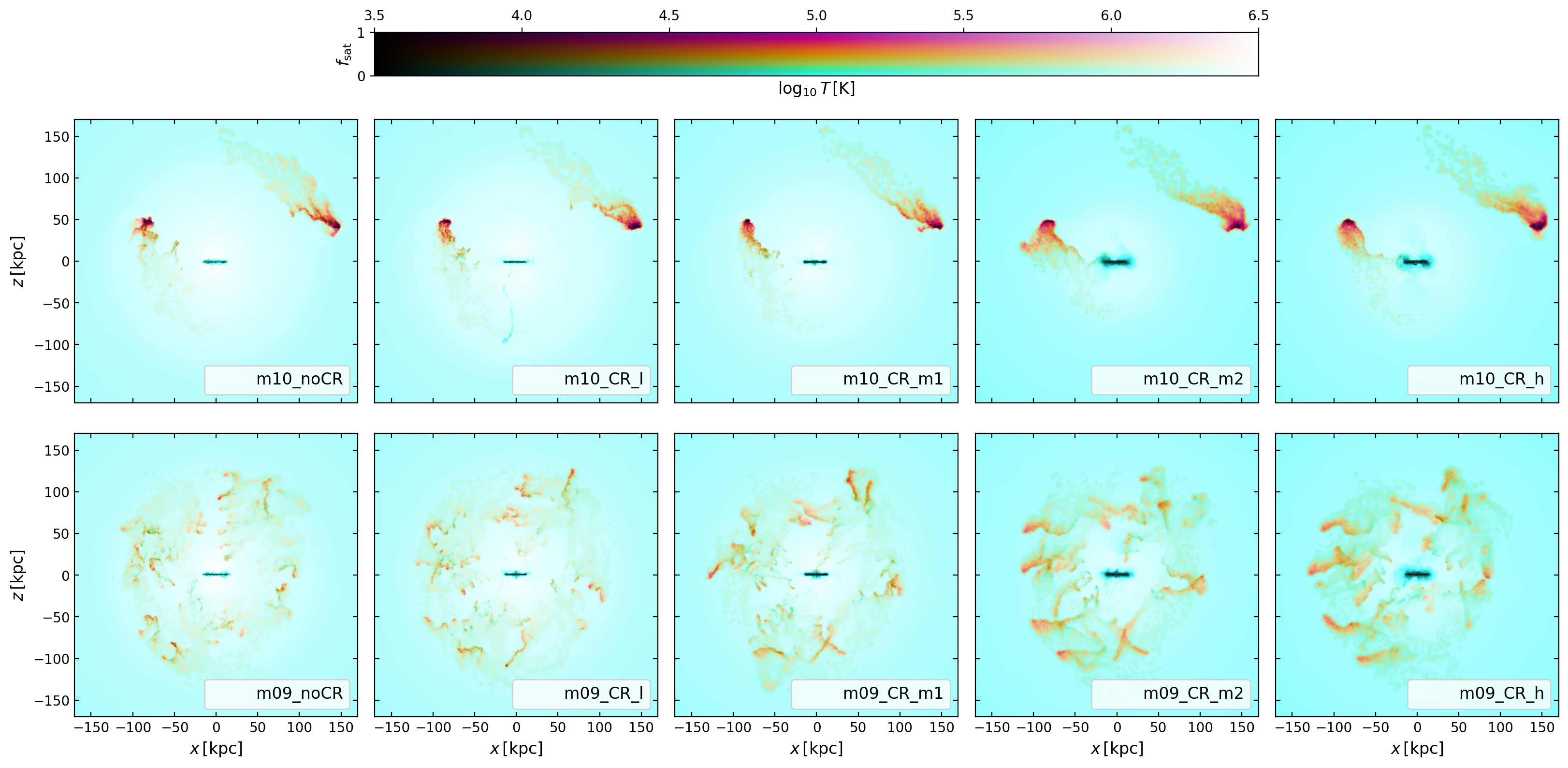}
\caption{The temperature distribution of different simulations (see Table \ref{t:runs}) at snapshot t = 1 Gyr. The colorbar varies both color and saturation based on $f_{\rm sat}$ and temperature, respectively. The parameter $f_{\rm sat}$ represents the local mass fraction that originated from the satellite, such that $f_{\rm sat}=1$ is entirely composed of satellite gas and $f_{\rm sat}=0$ is entirely host gas. The stripped cold gas is streaming behind the satellites and falling towards the central disk. There is also induced cool gas in the mixing layer of stripped cool gas and hot host gas. The top and bottom panels denote m10 and m09 runs, respectively. In each panel, there is an increasing amount of CRs from left to right, with the left-most being no-CR and the right-most being the highest amount of CRs. The ram-pressure-stripped clouds grow more and get bigger with an increase in the CR content (Left to Right).}
\label{f:snapshots}
\end{figure*} 
\subsection{How do Cosmic Rays affect Ram-pressure stripped cold clouds?}
In this section, we present an overview of the CGM gas evolution in the host galaxy using temperature snapshots at 1 Gyr from a suite of simulations. In Figure \ref{f:snapshots}, we show the edge-on projections of the temperature distribution for two satellite systems--\texttt{m10} (top panels) and \texttt{m09} (bottom panels)--under varying CR content. The CR energy density increases from left to right across the panels, with the left-most column representing the no-CR case and the right-most corresponding to the highest CR.

To generate these maps, we interpolate the gas particle data onto a uniform $512^3$ grid using a smoothing kernel based on each particle’s smoothing length, following a standard SPH-like deposition method. During deposition, temperature values are mass-weighted within each grid cell. For the final projection, we apply an $n^2$ weighting along the line of sight, roughly corresponding to a luminosity-weighted temperature map.

We differentiate between gas originating from satellites and that from the host using the satellite fraction parameter, $f_{\rm sat}$. This quantity is computed as a mass-weighted average along the projection axis, excluding gas associated with the host ISM and particles with temperatures exceeding $3\times10^5$\,K. A value of $f_{\rm sat} = 1$ indicates purely satellite-origin gas, while $f_{\rm sat} = 0$ denotes purely host-origin gas. In the visualizations, temperature is encoded through color saturation. All panels are centered on the host galaxy, and the thin blue strip at the center ($(0,0)$) represents the cold ISM of the host.

The similarity of this snapshot across the different runs indicates that all the simulations tell a similar general story. Initially, cold gas is primarily confined to satellites, with short trailing tails. Over time, these tails lengthen as cold gas escapes the satellite’s gravitational influence and moves toward the central galaxy. The increasing mixing with the host CGM is reflected by a color shift from red to orange. Cold gas is predominantly located near the central disk, in outflow regions, or along stripped tails. The satellite fraction $f_{\rm sat}$ decreases smoothly along these tails, indicating progressive mixing. Much of this cooled gas spatially coincides with the stripped cold gas and appears orange in color, suggesting significant mixing between the two components. Moreover, the stripped cold gas clouds are larger in simulations with the more massive m10 satellites compared to the lower-mass m09 systems. 

This visual comparison serves as an initial step in understanding the influence of CRs on the cold gas in the CGM.  With increasing CR content, the cold gas structures become more coherent and extended, exhibiting larger surface areas. This effect is particularly evident in the m09 simulations with smaller clouds: by $t = 1$ Gyr, the cold clouds in the no-CR runs are almost entirely disrupted and mixed with the CGM, whereas those in the high (\texttt{h}) and mid (\texttt{m2}) CR runs retain their structural integrity. A similar trend is observed for the m10 satellites: cold gas stripped from the nearest satellite is nearly dissolved in the no-CR case but remains intact in the high (\texttt{h}) and mid (\texttt{m2}) CR scenario. Even for the more distant m10 satellite, the presence of CRs leads to a visibly larger and more inflated ram-pressure stripped tail, while in the absence of CRs, the tail is composed of narrower filaments, 
suggesting enhanced destruction of cold gas. The snapshots also reveal more mixing-layer cooling (indicated by the orange color) occurring within the host CGM in the case of high-CR than no-CR, probably attributed to the larger surface area of the clouds. Therefore, these clouds grow more under the influence of CR. 

In summary, CRs appear to significantly influence the size and survival of ram-pressure stripped cold gas in the CGM. Higher CR content leads to larger, more coherent cold gas clouds that resist mixing and persist longer. 
In the following sections, we will quantitatively confirm our visual interpretation that CRs enhance mixing-layer cooling by increasing the cloud surface area.
\begin{figure*}
\includegraphics[width=0.49\linewidth]{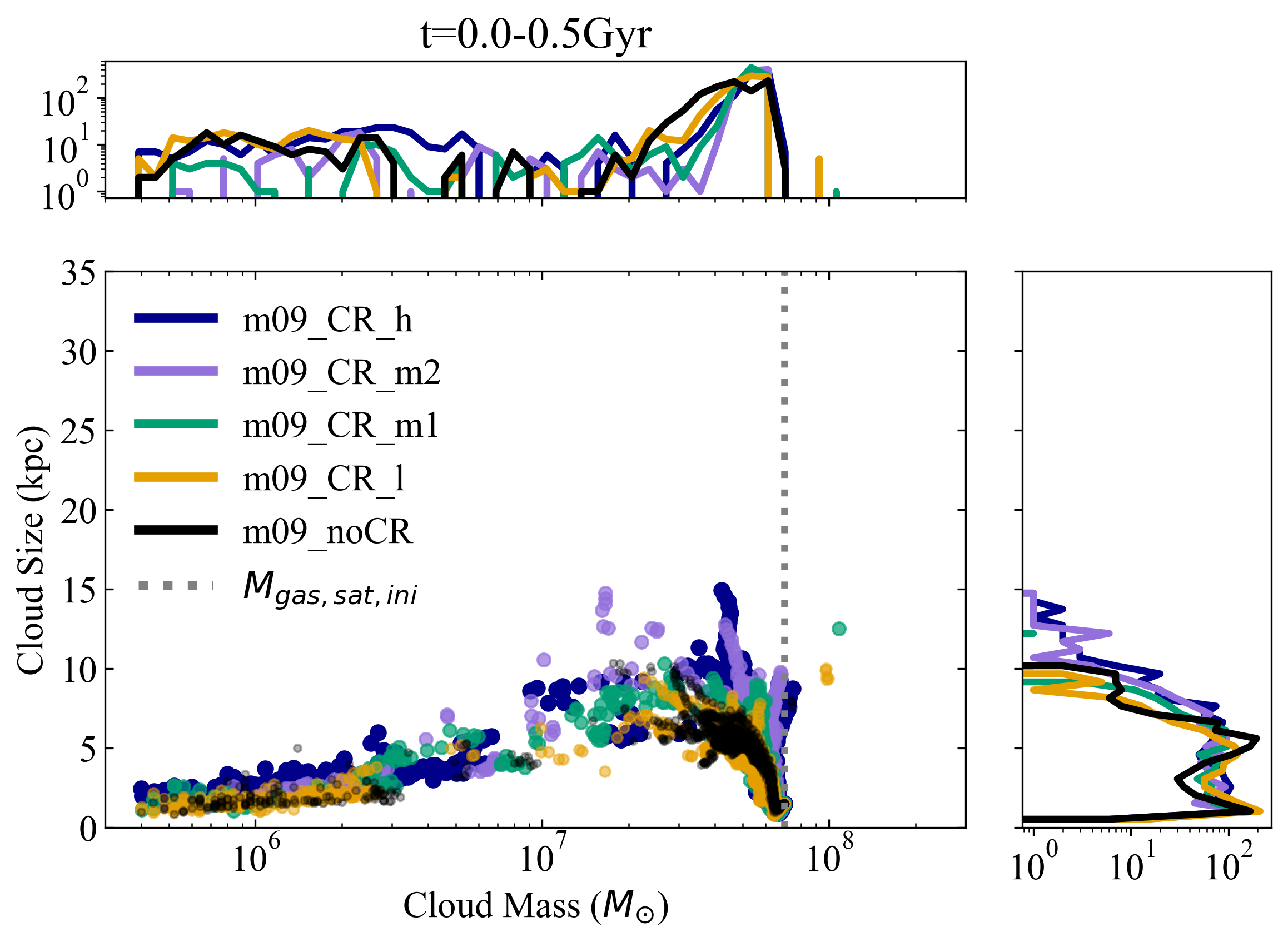}
\includegraphics[width=0.49\linewidth]{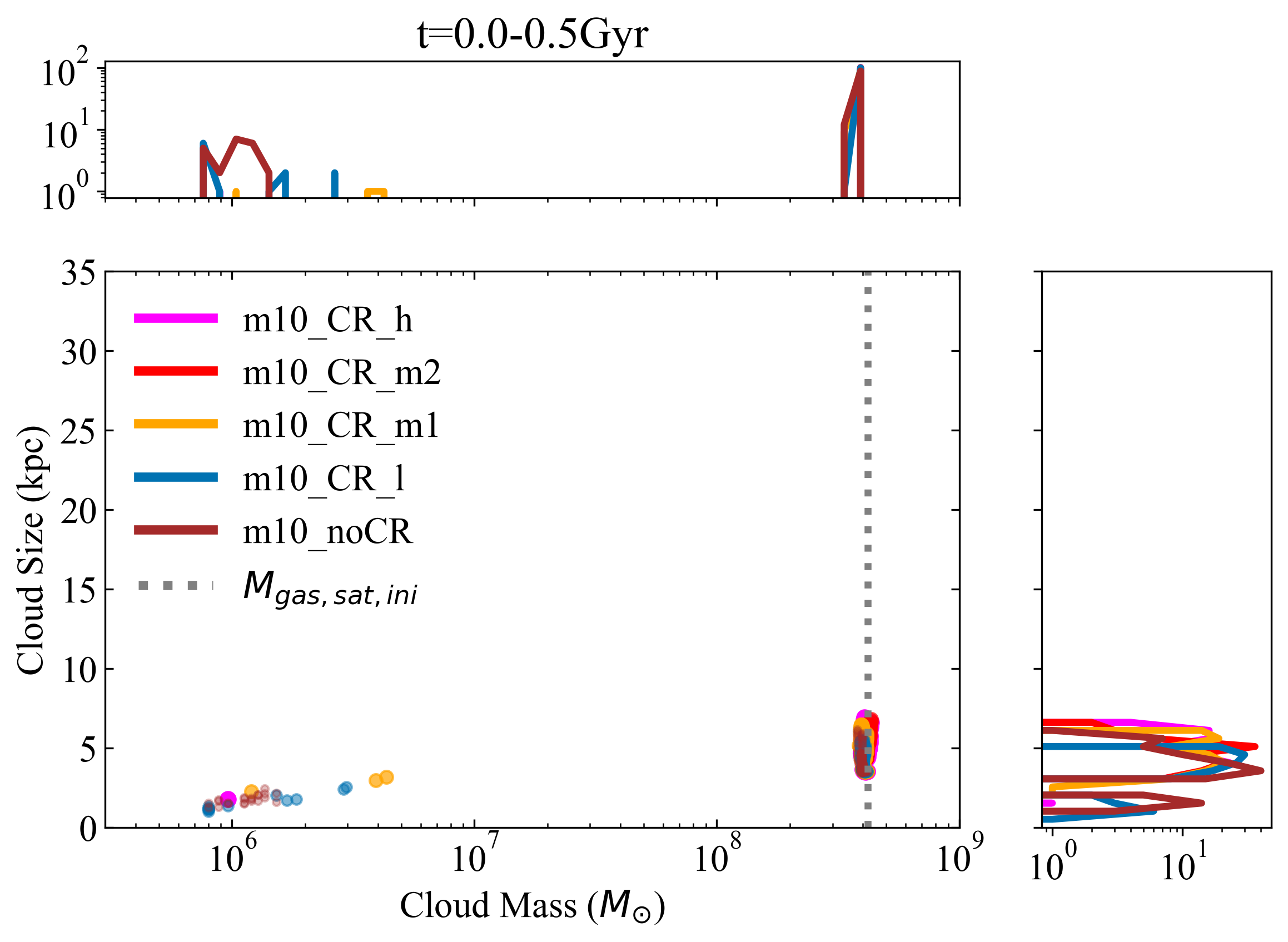}
\includegraphics[width=0.49\linewidth]{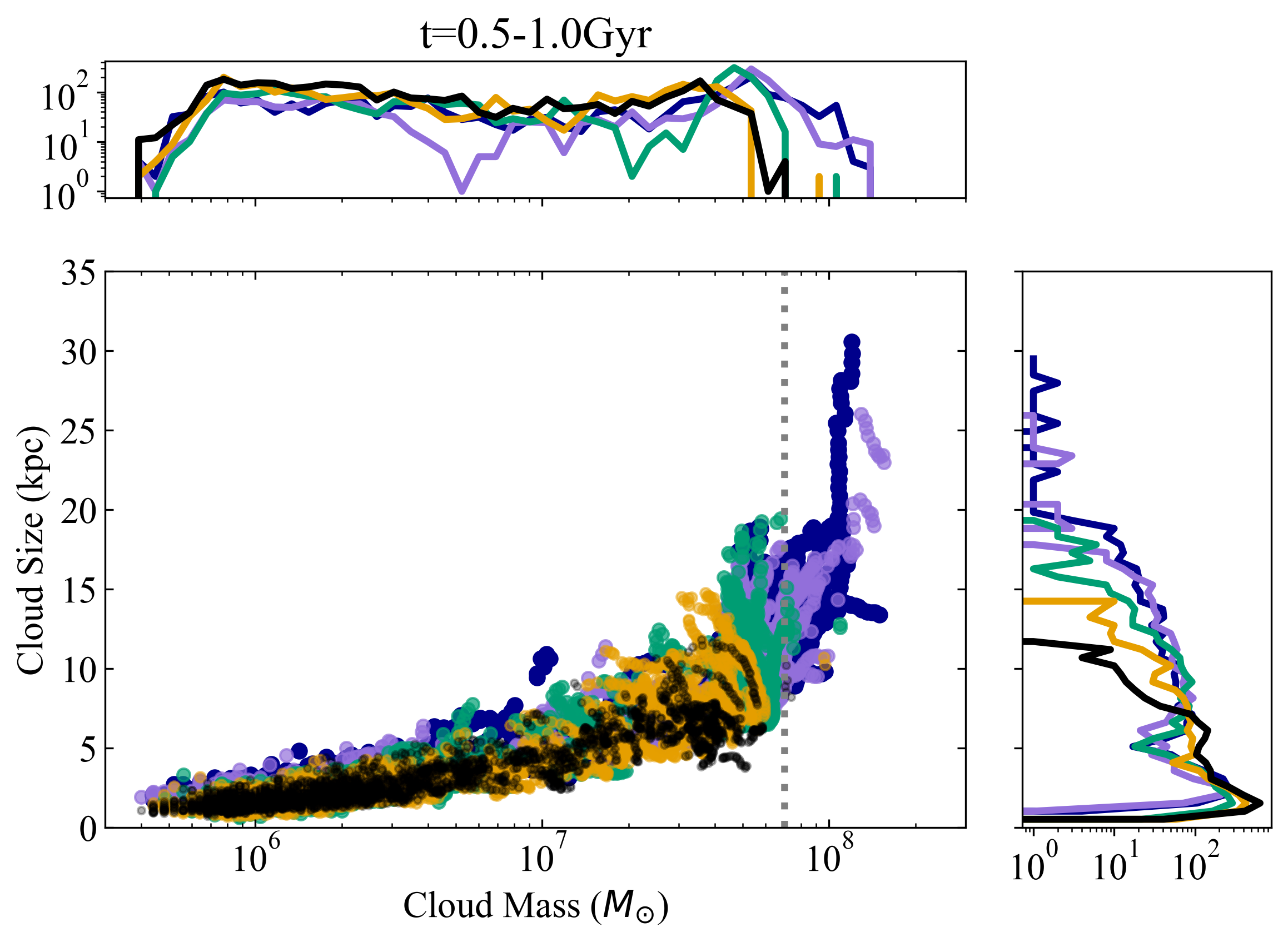}
\includegraphics[width=0.49\linewidth]{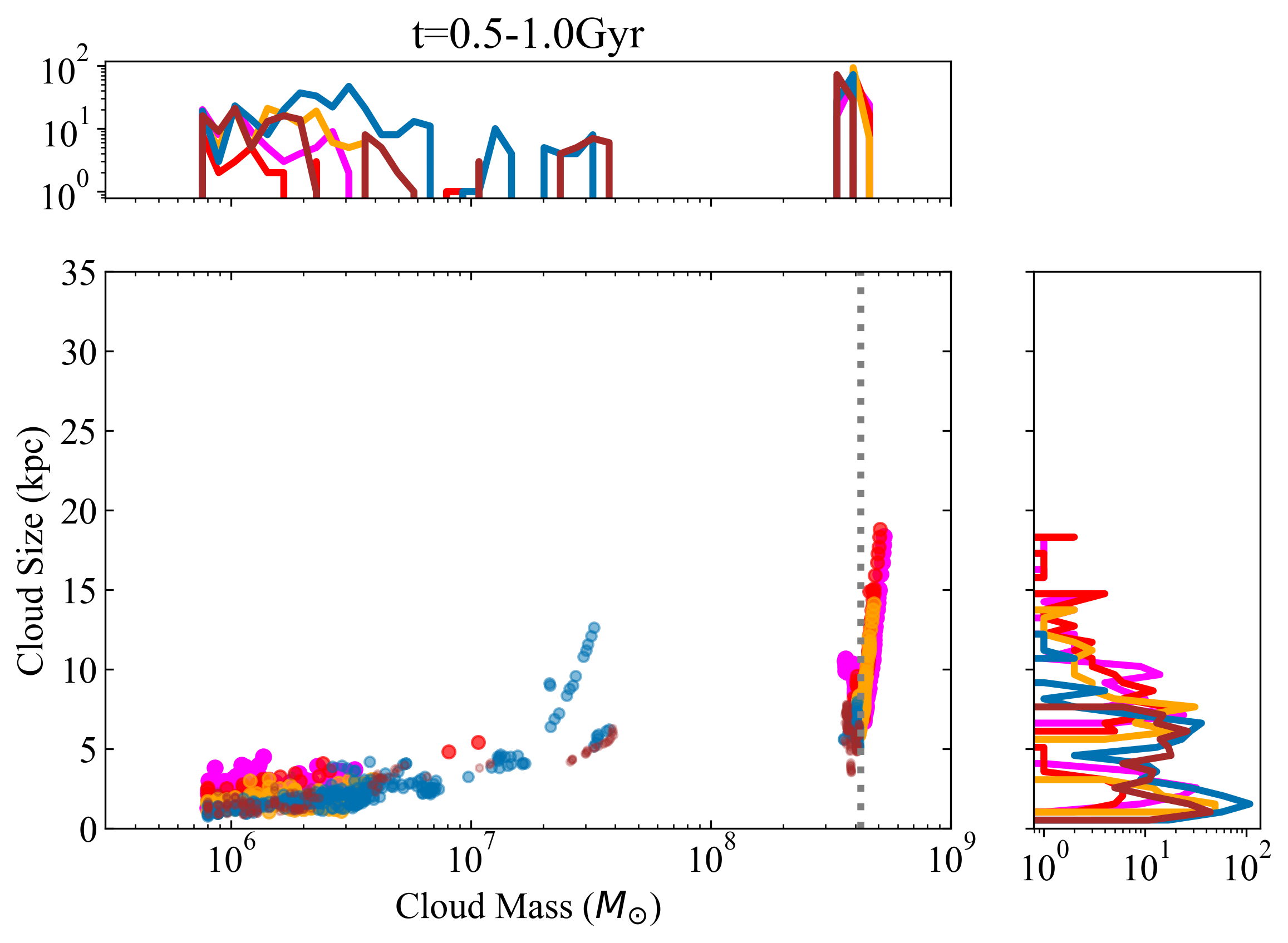}
\includegraphics[width=0.49\linewidth]{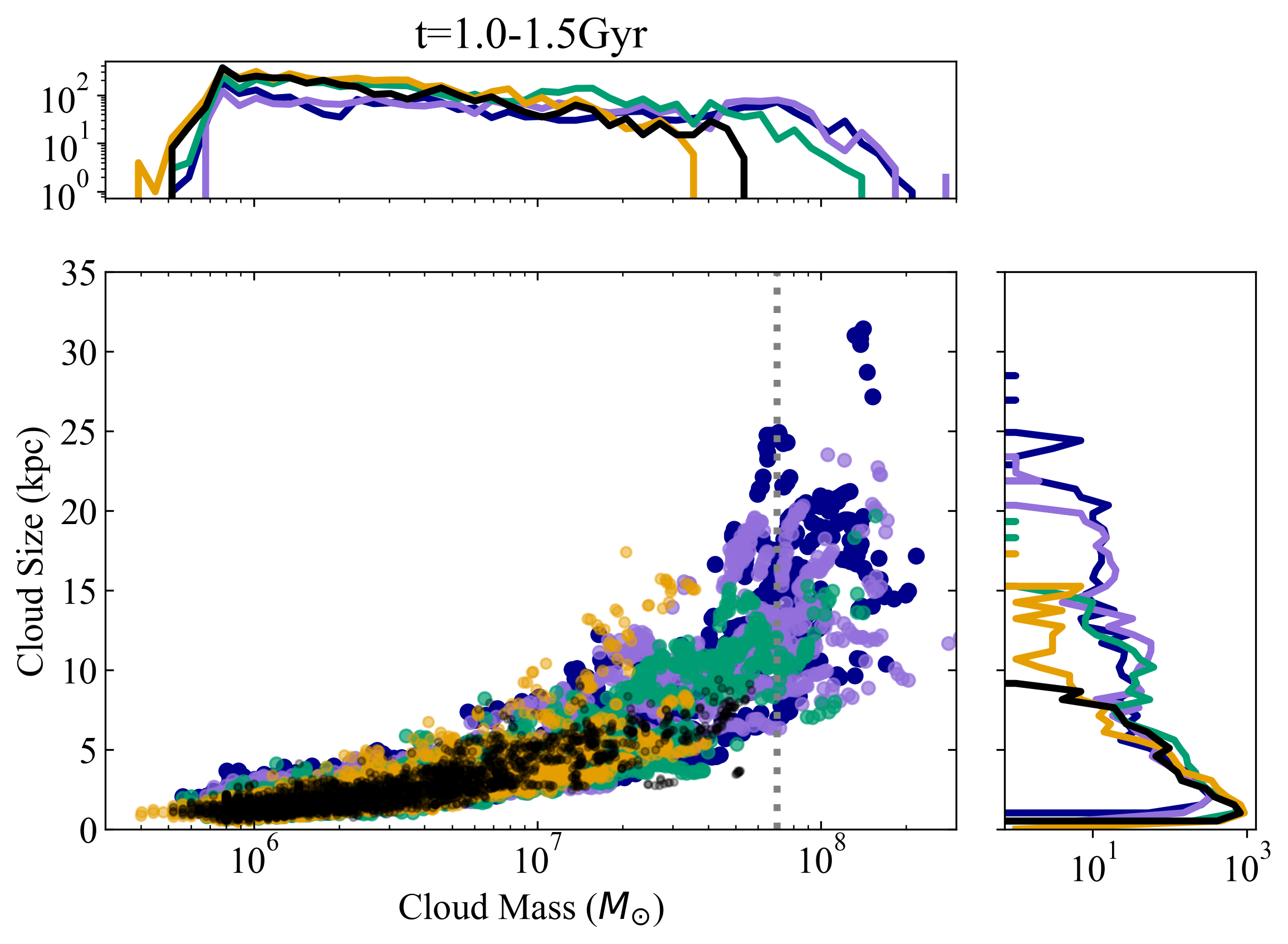}
\includegraphics[width=0.49\linewidth]{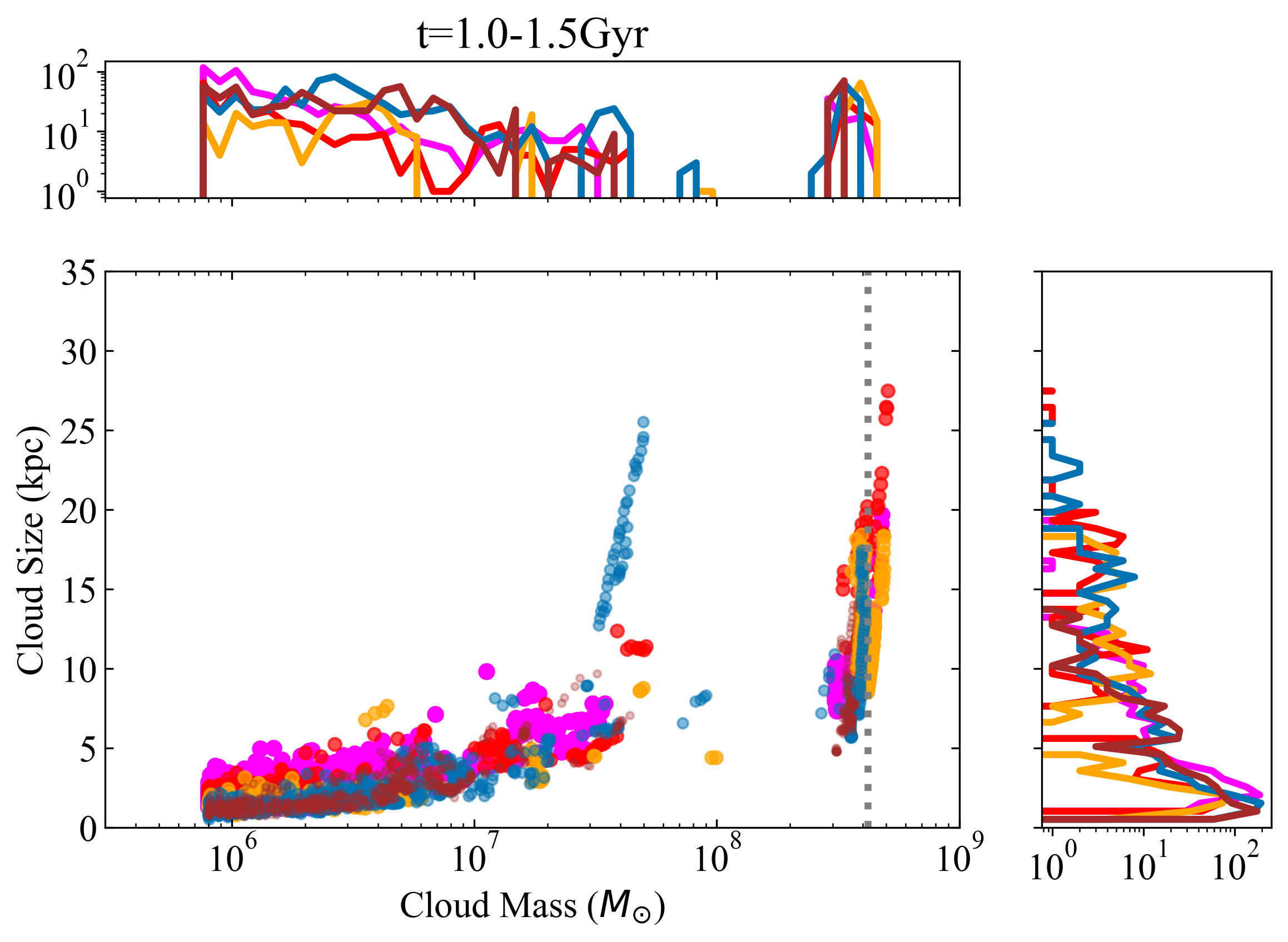}
\caption{Size and mass distribution of cold gas clouds identified using the Friends-of-Friends (FoF) algorithm across three time bins up to 1.5 Gyr (top to bottom rows). The left and right panels correspond to the m09 and m10 simulations, respectively. Each panel includes one-dimensional histograms along the top and right axes, showing the distributions of cloud mass and size, respectively. Cold clouds are defined using a temperature cut of $T < 3 \times 10^4$\,K  and identified using a linking length of 2\,kpc with a minimum group size of 10 particles. CRs enhance the formation of larger, more coherent cold gas clouds by suppressing fragmentation and enabling their long-term survival and growth through enhanced mixing-layer cooling.}
\label{f:cloud}
\end{figure*}
\subsection{Do CRs affect the size of the clouds?}
\label{S:size}
Motivated by the trends observed in the visual inspection, we proceed to quantitatively assess the mass and size of the cold gas clouds identified in each simulation. To this end, we apply a Friends-of-Friends (FoF) algorithm to the gas particles. Cold gas particles are defined using a temperature threshold of $T < 3 \times 10^4$\,K. 
For cloud identification, we adopt a linking length of 2\,kpc and require a minimum of 10 particles per group. Thus, any collection of at least 10 cold particles within a 2\,kpc linking length is identified as a coherent cold gas cloud. Additionally, we restrict our analysis to clouds located beyond 40\,kpc from the center of the host galaxy, as that is our definition of the edge of the ISM in Section \ref{S:Def} (this means the central position of the cloud is beyond 40 kpc). These specific criteria of linking length and minimum particle number are chosen to ensure that the total mass of the identified clouds matches the total mass of cold gas particles ($T < 3 \times 10^4$\,K) beyond 40\,kpc, thereby providing a complete and consistent accounting of the cold gas in the outer CGM.

In Figure~\ref{f:cloud}, we present the distribution of cloud 
radii and masses across three time bins up to 1.5 Gyr, with left and right panels corresponding to m09 and m10 runs, respectively. For each panel, we also include 1-D histograms along the top and right axes, representing the mass and radius distributions, respectively. The plotted data include all cold clouds identified in each snapshot within a given time bin (50 snapshots per time bin). As a result, some clouds appear multiple times in a panel because they persist across multiple snapshots.

In the first time bin, the majority of identified cold clouds are associated with the satellites, based on their consistency in both mass and size with the satellite properties. For reference, the m09 satellite has a gas mass of $7 \times 10^7\,M_\odot$ and a gas disk radius of 0.87\,kpc, while the m10 satellite has a gas mass of $4.2 \times 10^8\,M_\odot$ and a gas disk radius of 2.1\,kpc. In the m10 cases, the clouds are more clearly associated with the satellites, as not much gas has been stripped by 0.5 Gyr.  However, we see a smoother distribution of cloud masses and sizes in the m09 runs, which is likely due to the earlier onset and more rapid stripping in those runs.  


Initially, there is no significant difference between the no-CR and CR runs.
At later times (the lower two rows), stripping and fragmentation become more prominent, with an increasing number of lower-mass and smaller clouds in all of our runs. In the intermediate time bin (particularly for m09 runs and slightly for large clouds  $>10^8\,M_{\odot}$ of m10 runs), we observe a monotonic trend in cloud size with increasing CR content- higher CR levels produce systematically larger clouds. The largest clouds in the high-CR runs exceed those in the no-CR runs by about 10-15\,kpc in radius. Additionally, the number of fragmented, low-mass clouds is also slightly lower in high (\texttt{h}) and mid (\texttt{m2}) CR simulations compared to other cases. This effect is more visible in the m09 run. 

CRs provide additional pressure support to the cloud, making it puffier and less dense. In the high-CR run, the cold gas fraction in the CGM with density less than $10^{-3}$/cc is $\sim24\%$, whereas the same is $0.3\%$ in no-CR. Extra pressure by CR can puff up the stripped gas. Also, CR energy is less vulnerable to radiative cooling losses, unlike thermal energy. Therefore, when CR pressure dominates, the reduction in thermal pressure due to cooling does not result in significant gas compression. Instead, the gas cools isochorically (at constant volume), as opposed to the typical isobaric (constant pressure) cooling in a thermally unstable gas, which leads to density enhancements. As a result, high CR pressure tends to produce larger, less dense cold clouds \citep{Butsky2020}.

In terms of mass, the most massive clouds in the no-CR and low-to-mid (\texttt{m1}) CR runs do not exceed the original gas mass of the satellites (Shown by vertical grey dotted line). In contrast, the high and mid (\texttt{m2})-CR runs produce clouds with masses roughly $\sim 2$ times that of the initial satellite gas mass. This growth can be attributed to the increased surface area of larger clouds, which enhances mixing with the ambient medium. The mixing layers facilitate cooling, allowing the clouds to grow further and persist for longer durations. Note that the radii of the clouds also increases by a factor of two, which also points directly towards the lower density (M/r$^3$) with CRs than the no-CR case. 

In the last time bin, the physical processes we describe above continue to lead to different cloud populations in the no-CR versus CR runs.  The clouds in the no-CR run tend to have lower masses and smaller sizes than at earlier times, indicating that they are highly fragmented, likely due to significant disruption and mixing with the ambient medium.
In contrast, the CR runs retain larger and more coherent cloud structures, with the high-CR case preserving the most intact and extended cold gas clouds. This highlights the role of CR pressure in stabilizing stripped cold gas against fragmentation over longer timescales.

In summary, CRs promote larger and more coherent cloud structures, suppressing fragmentation and enhancing survival. This behavior is due to CR pressure preventing isobaric collapse by enabling isochoric cooling. Consequently, high-CR runs form puffier, less dense clouds that grow and persist via enhanced mixing-layer cooling due to their larger surface area. By the final time bin, high-CR clouds remain intact, while no-CR clouds are heavily fragmented.

\subsection{Do Cosmic Rays affect the cold gas mass contribution from Satellites?}
\label{S:mass}
\begin{figure*}
\includegraphics[width=0.49\textwidth]{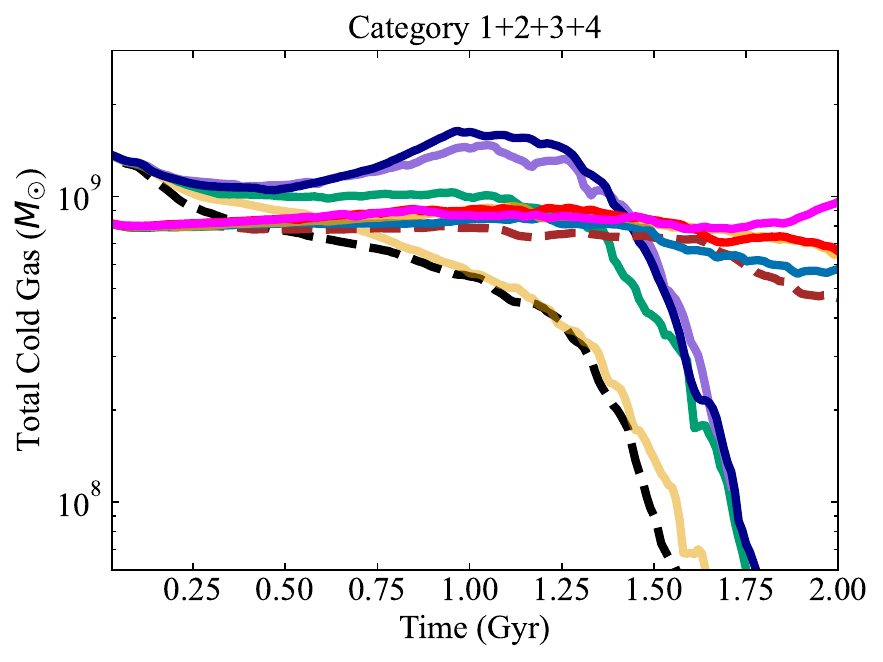}
\includegraphics[width=0.49\textwidth]{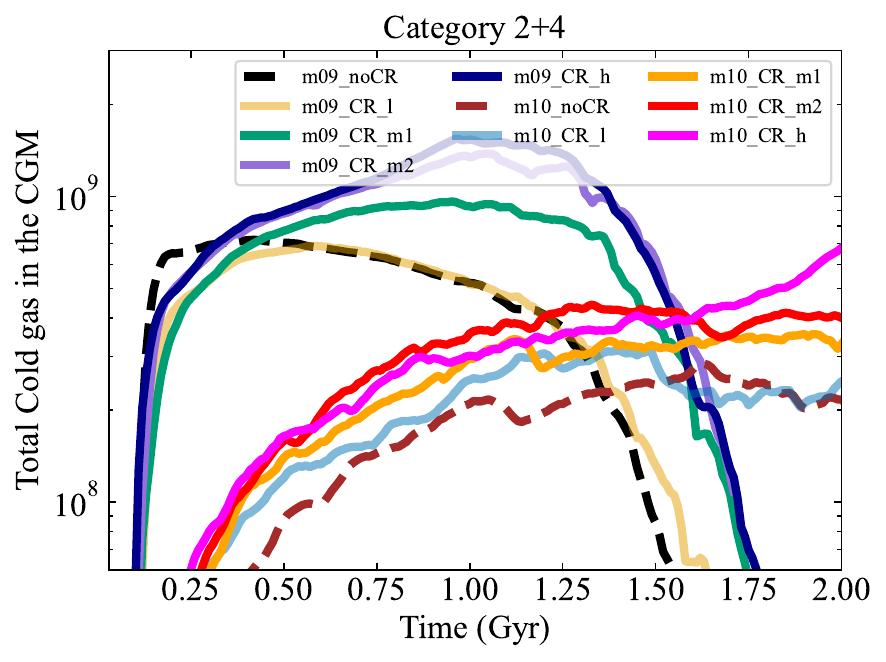}
\includegraphics[width=0.49\textwidth]{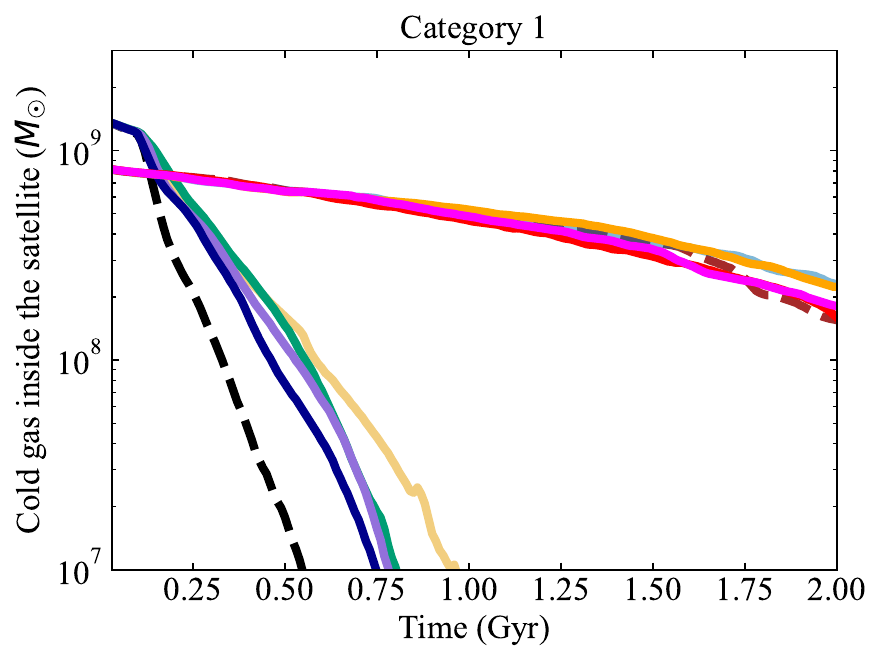}
\includegraphics[width=0.49\textwidth]{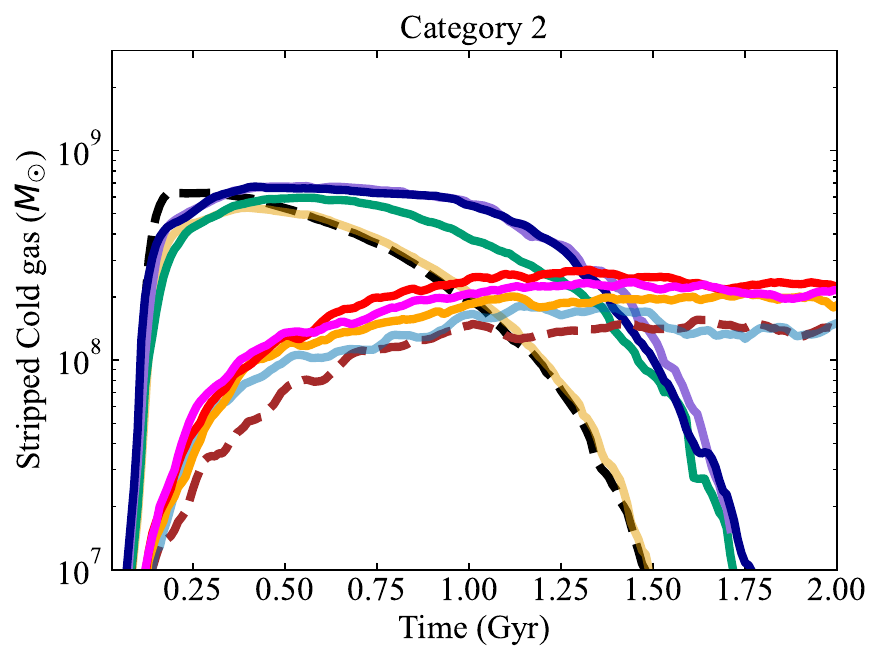}
\includegraphics[width=0.49\textwidth]{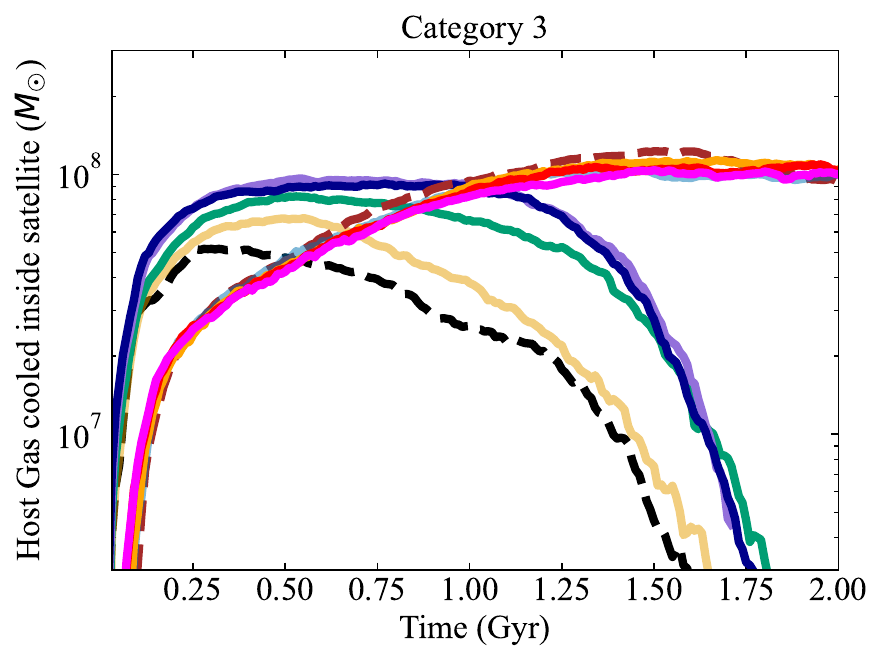}
\includegraphics[width=0.49\textwidth]{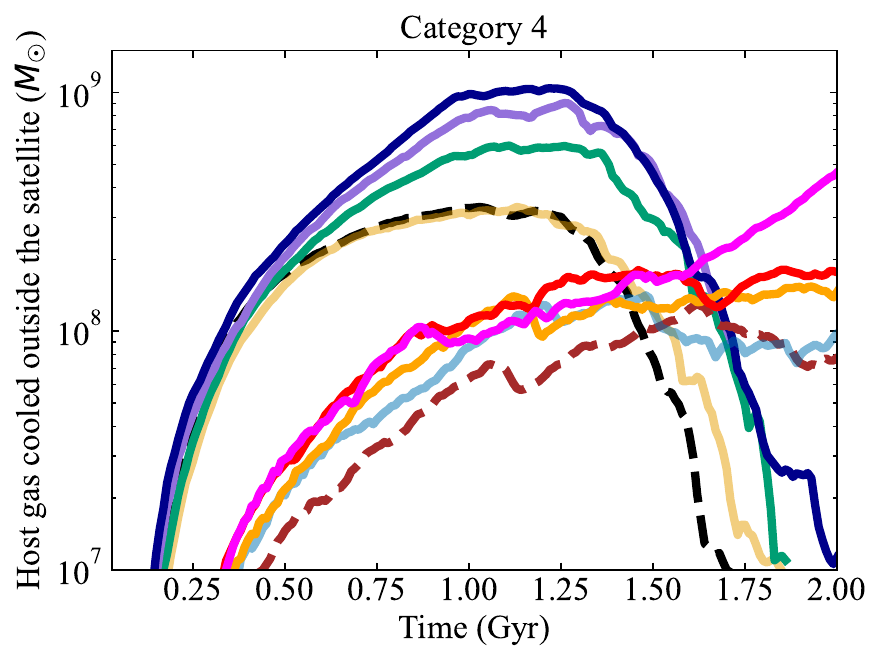}
\caption{Time evolution of cold gas mass ($T \leq 3\times10^4$ K) beyond 40 kpc from the host galaxy center, split into the four categories defined in Section \ref{S:Def}: cold gas inside satellites (1), stripped cold gas (2), host gas cooled inside satellites (3), and host gas cooled outside satellites (4). \textit{Top panel:} Total cold gas in the CGM and satellites (left; sum of middle and bottom panels) and total cold gas in the CGM only, excluding satellite-bound gas (right; sum of right panels in middle and bottom rows). \textit{Middle panel:}  Cold gas inside satellites (left) and stripped from satellites (right; defined as cold satellite gas beyond six satellite scale radii). \textit{Bottom panel:} Host cold gas cooled inside satellites (left) and outside satellites (right). An increase in CR pressure enhances total stripped cold gas mass, mixing layer cooling along with the total cold gas mass, with the largest difference between no-CR and high (\texttt{h}) and mid (\texttt{m2}) CR runs.}
\label{f:time_evo}
\end{figure*}

A natural follow-up question is whether the large, stripped cold gas clouds contribute significantly to the overall cold gas budget in the CGM. In this section, we examine whether CRs influence the total cold gas mass in the CGM that originates from satellite galaxies. Figure~\ref{f:time_evo} shows the time evolution of cold gas mass beyond 40\,kpc. The top left panel includes all cold gas outside this radius: gas stripped from satellites, cold gas residing within satellites, and cold gas formed via induced cooling of host halo gas (corresponding to the sum of middle and bottom panels in Figure~\ref{f:time_evo}. The right panel, by contrast, excludes cold gas inside satellites, accounting only for stripped material and host gas that has cooled outside the satellite boundary (summed from the right-middle and right-bottom panels in Figure~\ref{f:time_evo}).

For the m09 run, we observe a clear trend of increasing total cold gas mass with increasing CR pressure (Top panel). However, the difference between the no-CR and low-CR runs is relatively modest. In contrast, the m10 simulations exhibit minimal differences between CR and no-CR runs at late times, and the total cold gas mass appears largely insensitive to the level of CR pressure among the CR runs. A similar trend holds for the total CGM cold gas mass. Notably, around 1\,Gyr, the total cold gas mass in the CGM for the high-CR case is $\sim4$ times greater than that in the no-CR run.

To better understand the contribution of each component to the cold gas budget, we now investigate which component (classification in different categories is shown in Section \ref{S:ic}) is most affected by the presence of CRs. The middle-left panel of Figure~\ref{f:time_evo} shows the evolution of cold gas inside the satellites (Category 1), revealing that the stripping timescale is generally longer in CR runs compared to the no-CR case. Interestingly, however, we observe a non-monotonic trend with CR fraction. In the case of m09, the high-CR run exhibits the most rapid stripping (within $\lesssim 1$\,Gyr), followed by the mid-CR run ($\sim1$\,Gyr), and the low-CR run, which shows the slowest stripping rate (extending beyond 1\,Gyr). However, these differences in mass are low and hence not significant enough considering the mass resolution of our simulation. Despite the faster stripping in the high-CR case, the stripped gas (Category 2) remains in the CGM for a longer duration, as seen in the sustained cold gas mass in the middle-right panel of Figure~\ref{f:time_evo}. This persistence may indicate longer cloud survival time, as discussed further in Section~3.5.

The bottom two panels of Figure~\ref{f:time_evo} display the evolution of induced cooling in the host CGM, distinguishing between gas cooled inside (bottom-left; Category 3) and outside (bottom-right; Category 4) of the satellite boundaries. In both cases, we find a clear trend: stronger CR pressure leads to enhanced cooling of the host gas. The cooling of the host gas inside satellites increases monotonically with CR strength in the m09 run. Similarly, for gas that cools outside the satellite--primarily due to mixing layer processes--the high-CR and mid-CR runs show significantly more cold gas than the low-CR and no-CR cases. This enhanced mixing-layer cooling contributes substantially to the total cold gas content in the CGM, reinforcing the role of CRs in promoting cold gas growth beyond 40\,kpc.
The cold gas mass trends in the m10 runs are less pronounced but follow similar trends to the m09 runs.

In summary, CRs enhance the total cold gas mass in the CGM by affecting the contributions of stripped gas (middle-right panel of Figure~\ref{f:time_evo}) and mixing layer cooled gas (lower-right panel of Figure~\ref{f:time_evo}). Higher CR pressure increases the longevity of stripped gas in the CGM. CRs also promote greater induced cooling of the host gas in the mixing layer. These effects collectively lead to a sustained and elevated cold gas reservoir in the CGM with increasing CR strength.

\subsection{Do CRs Affect the Covering Fraction of Cold Gas?}
\label{S:cv}
In this section, we examine the covering fraction of cold gas--an observable quantity that provides insight into the spatial distribution of the CGM. 
Given that CRs tend to produce larger and more coherent cold gas clouds, we expect a corresponding increase in covering fraction with increasing CR content.

Since it is observationally challenging to distinguish between satellite-origin and host-origin gas, we compute the covering fraction by considering all cold gas particles with $T < 3 \times 10^4$\,K located beyond 40\,kpc, irrespective of origin. We calculate covering fractions for two viewing geometries: (1) an external observer viewing the galaxy from afar, and (2) an internal observer situated at the center of a Milky Way-like galaxy (MW). For the MW case, we shoot 1000 sightlines radially outward from the galactic center, covering angles from $-180^\circ$ to $180^\circ$. For the external view, we use 1000 parallel sightlines passing through the CGM (40 kpc $<$ r $<$ 200 kpc).

\begin{figure*}
\includegraphics[width=\textwidth]{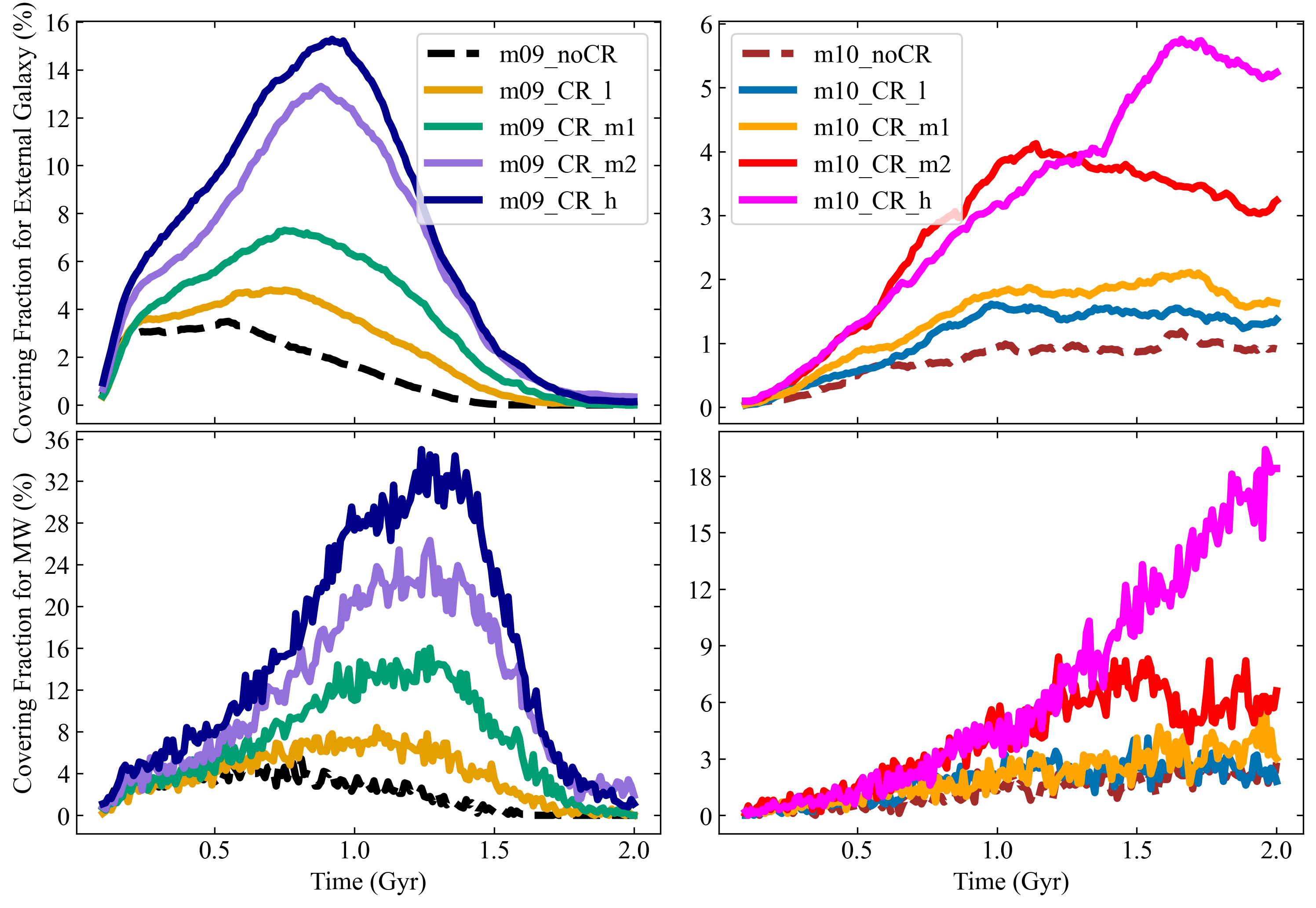}
\caption{Covering fraction of cold gas ($T < 3 \times 10^4$\,K) as a function of time for different CR strengths for the m10 (right panel) and m09 (left panel) runs. The bottom panel shows the covering fraction for a Milky Way (MW) view, with sightlines originating from the galactic center, while the top panel represents an external galaxy view using parallel sightlines. In both geometries, the covering fraction increases with CR content, with the high-CR runs showing the most significant enhancement, up to $\sim$2--8 times higher compared to no-CR.}  
\label{f:CF}
\end{figure*}

Figure~\ref{f:CF} shows the resulting covering fractions of cold gas for both geometries, with the top and bottom panels representing the external galaxy and MW views, respectively, for m09 (left panel) and m10 (right panel) runs. We note that the covering fraction in the MW view is higher than in the external galaxy view. This is primarily due to the radial nature of the sightlines. 
Additionally, the m09 runs, which contain a larger number of satellites compared to the m10 runs, exhibit higher covering fractions overall, as the increased number of stripped structures leads to broader spatial coverage in the CGM.
The high-CR cases exhibit significantly higher covering fractions. In the MW view, high-CR runs show up to $\sim5$--$7$ times greater covering fraction compared to the no-CR case. A monotonic increase in covering fraction is observed with increasing CR strength, although we highlight that the change is particularly dramatic only in the two highest-CR scenarios. 


In the external galaxy view, the covering fraction increases steadily with CR strength. In the m09 run, the covering fraction rises from $\sim2\%$ in the no-CR case to as high as $\sim17\%$ in the high-CR run - an eightfold increase. 
A similar trend holds for the m10 run, though the overall covering fractions are lower. Again, the increase is most pronounced in the high-CR scenario, while differences between the no-, low(l)-, and mid(m1) -CR runs remain modest.


Observations of the cold phase of the CGM, both in the Milky Way and in external galaxies, have revealed high covering fractions. For instance, \citet{Shull2009} reports a covering fraction of $81 \pm 5\%$ for SiIII high-velocity clouds (HVCs) across high-latitude Galactic sightlines. Similarly, \citet{Lehner2012} find covering fractions of 50\% and $\leq 67\%$ for intermediate-velocity HVCs (iHVCs) in stellar and AGN sightlines, respectively. In the case of external $L^*$ galaxies, \citet{Werk2013} find a $\sim70\%$ covering fraction for MgII within 75 kpc, and \citet{Thom2012}, \citet{Tumlinson2013}, \citet{Werk2013} and \cite{Wilde} report that the total covering fraction of cool gas reaches up to $80-90\%$ for HI column densities $N_{\rm HI} > 10^{14}$\,cm$^{-2}$. While it comes as no surprise that our idealized insertion of a small number of satellites has much lower covering fractions than observations, our results suggest that CRs can play a significant role in producing such high covering fractions. By promoting the formation of larger, more coherent cold gas clouds and suppressing fragmentation, CRs enhance the spatial extent and survival of the cold phase. This leads to a systematically higher covering fraction in our high-CR simulations compared to runs without CRs. Therefore, CR-driven cloud growth and stabilization may provide an 
important physical mechanism allowing for the high covering fractions observed in both the Milky Way and external galaxies.

In summary, CRs enhance the covering fraction of cold gas in the CGM, with the effect being most pronounced in the high-CR runs. In the high-CR m09 case, the covering fraction increases by up to a factor of 8 compared to the no-CR scenario, underscoring the role of CRs in boosting the spatial visibility of cold gas. Hence, CRs can be one possible way to explain the observed high covering fraction of cold gas. This effect is especially evident in the Milky Way view due to radial sightlines, while also present in external galaxy projections.  


\begin{figure*}
\includegraphics[width=
0.5\linewidth]{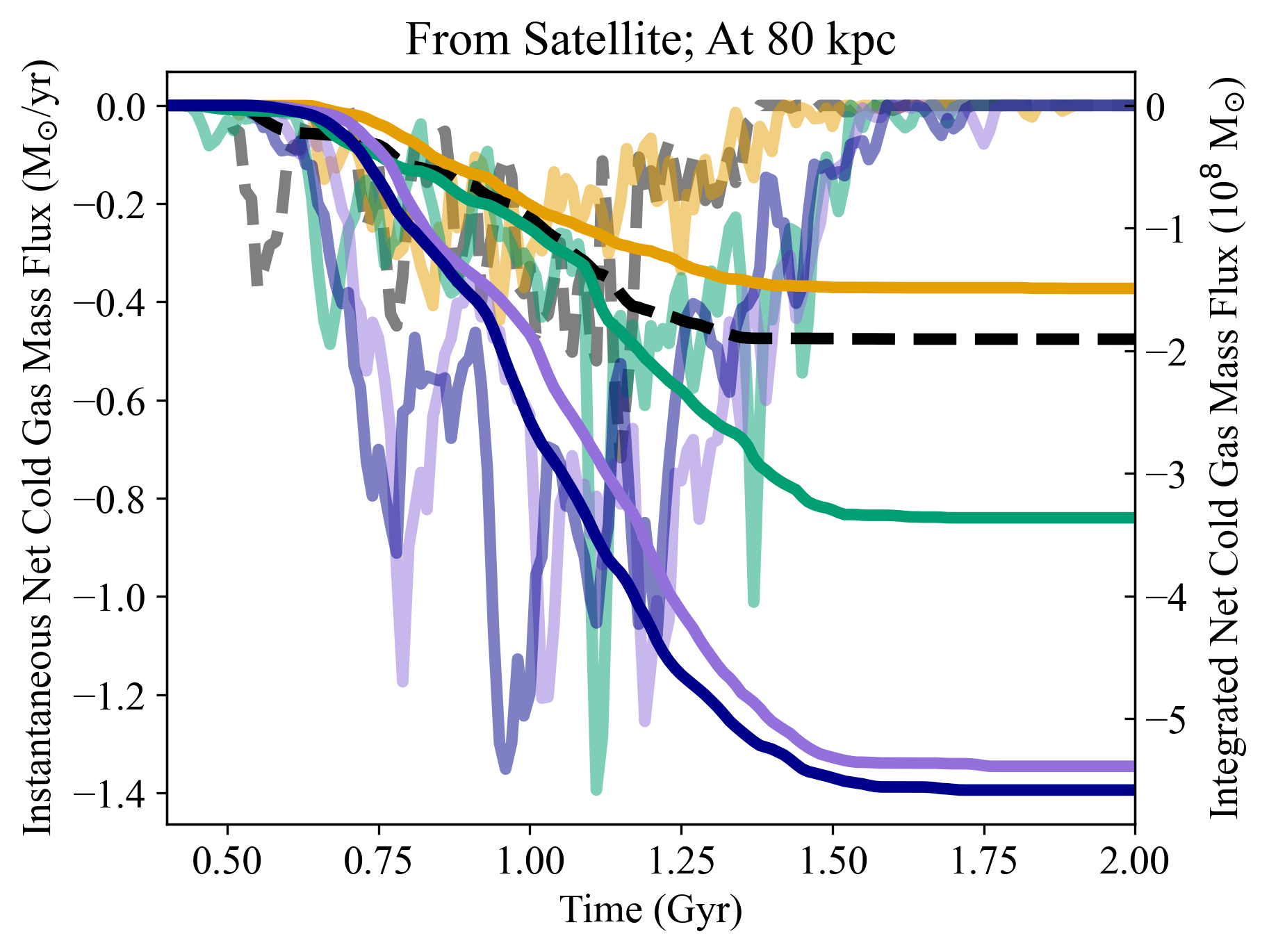}
\includegraphics[width=0.5\linewidth]{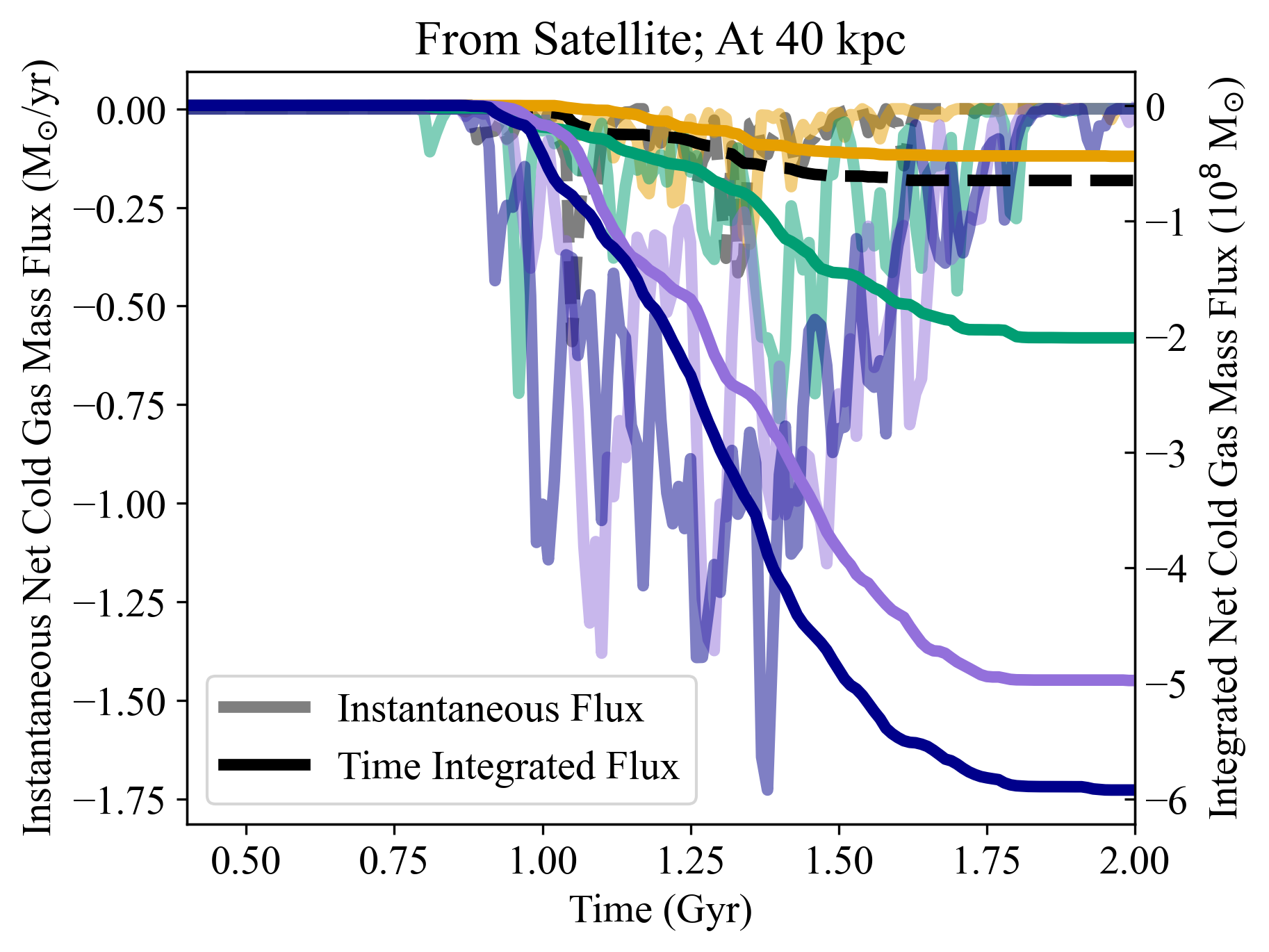}
\includegraphics[width=0.5\linewidth]{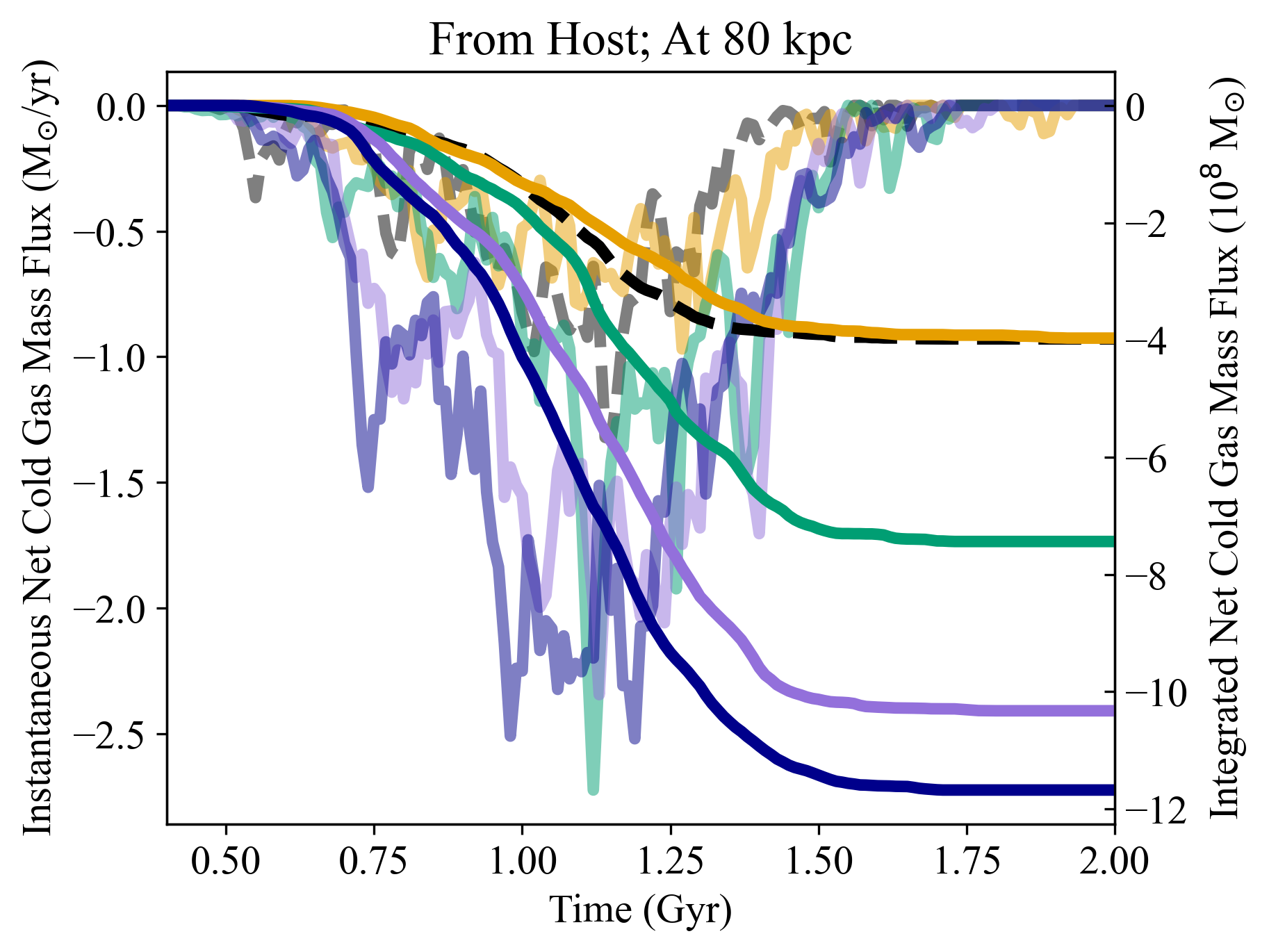}
\includegraphics[width=0.5\linewidth]{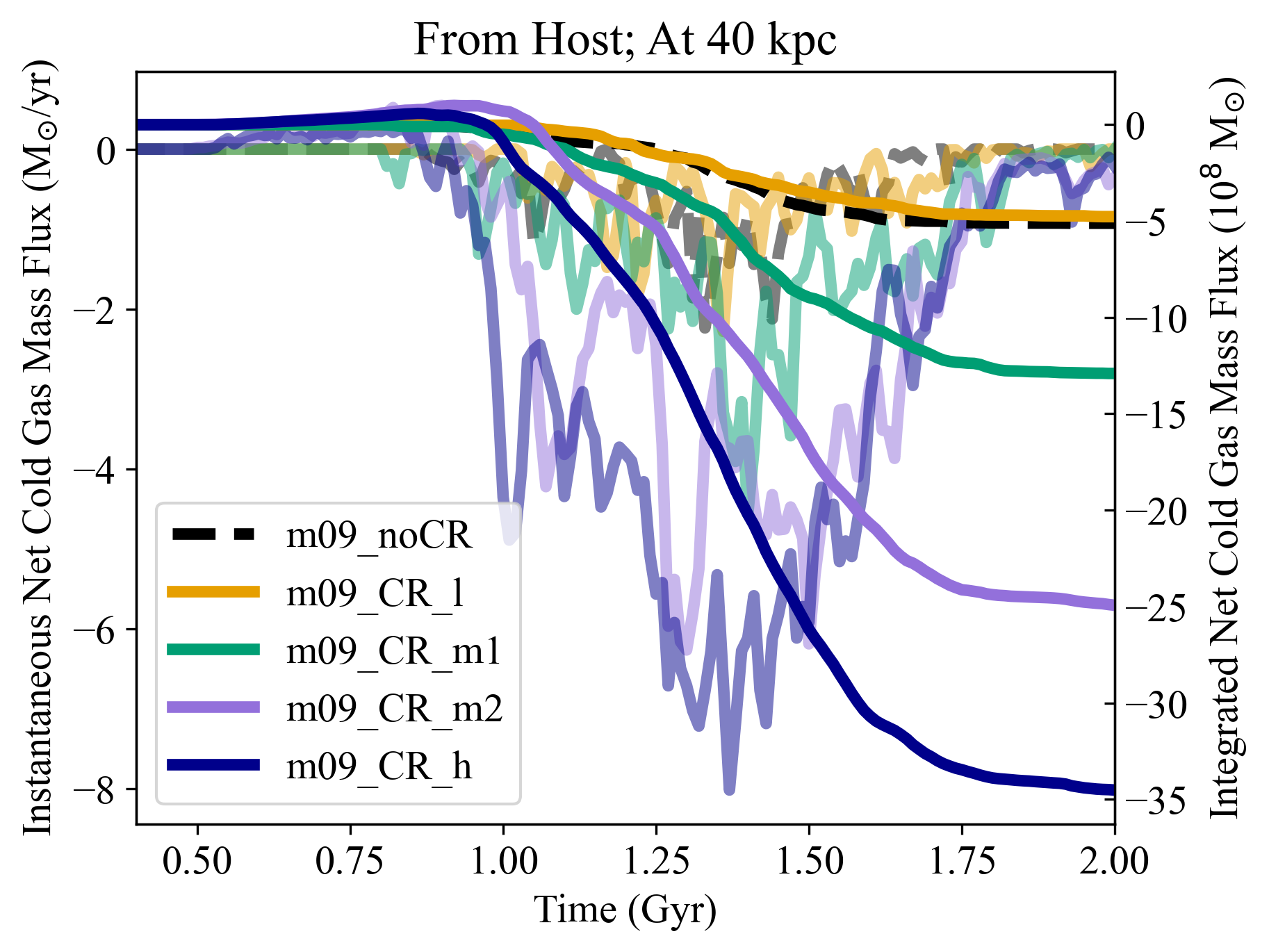}
\caption{The time evolution of the net cold (T$\leq3\times10^4$K) gas mass flux (instantaneous by light-shaded and time-integrated by dark-shaded) at $40$ kpc (\textit{Right}) and $80$kpc (\textit{Left}) shell of $1$ kpc width for the m09 case (where negative indicates inflow). The top and bottom panels indicate the mass flux from satellite gas and host gas, respectively. Increasing CR content enhances the inflow of cold gas into the central galaxy, with the high-CR run exhibiting an inflow rate nearly $\sim2-4$ times that of the no-CR case. }
\label{f:mass_flux}
\end{figure*}

\begin{figure*}

\includegraphics[width=0.49\linewidth]{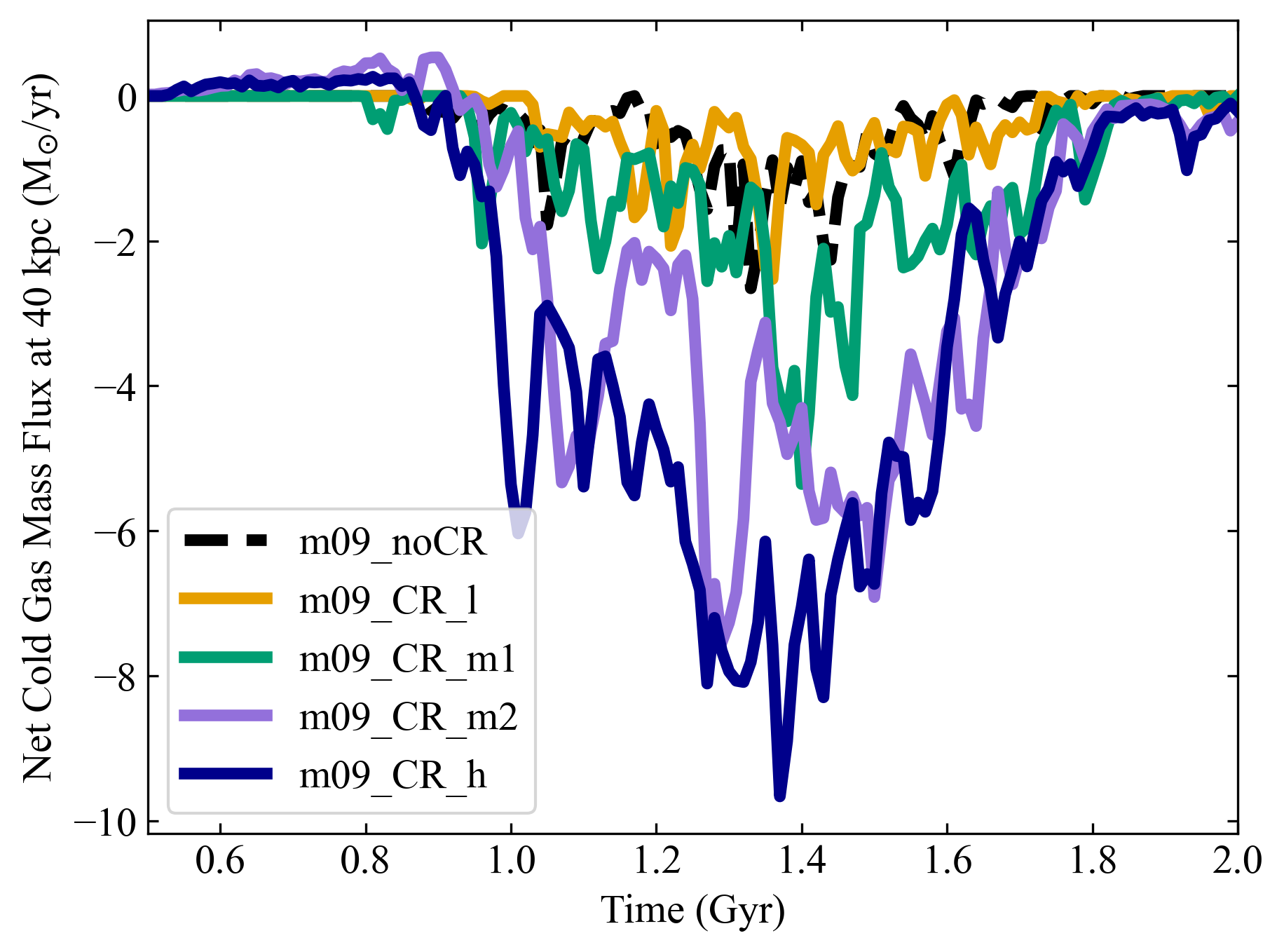}
\includegraphics[width=0.49\linewidth]{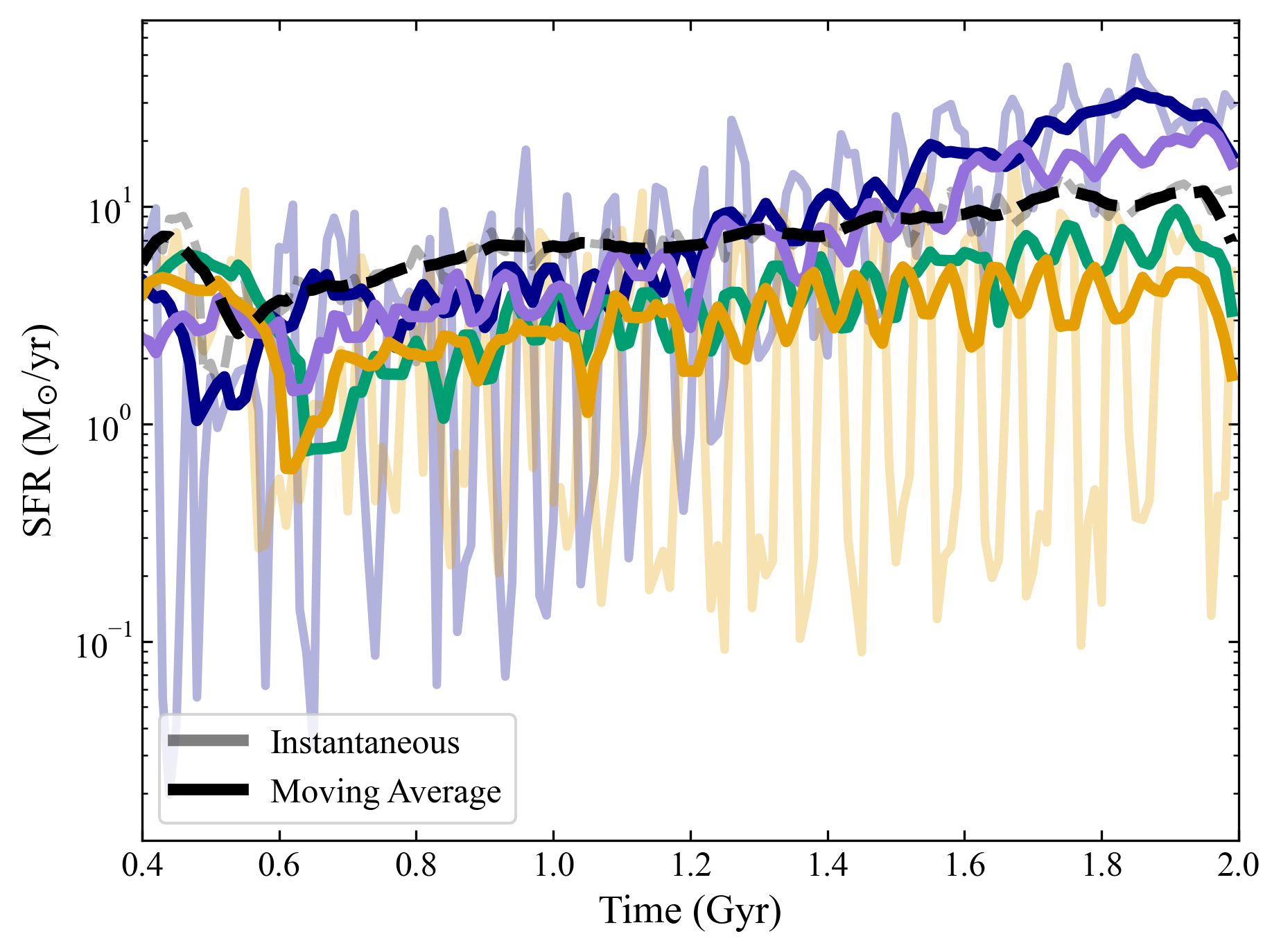}
\caption{Left: Total net flux of cold gas (where negative indicates inflow) at 40 kpc for the m09 simulations, obtained as the sum of the net flux components shown in the top and bottom right panels of Figure \ref{f:mass_flux}.
Right: SFR of the host galaxy for different CR runs. Light-shaded curves show the raw SFR time series (we only show them for no-CR, low-CR and high-CR in order to avoid a crowded plot), while bold curves indicate smoothed trends (moving average). CR runs display stronger variability in SFR and can reach up to twice the SFR of the no-CR case. In high-CR runs, the burstiness declines later due to a sustained influx of cold gas unlike in low-CR runs. While high-CR runs have similar average SFRs to the no-CR case at early times, their SFRs rise to roughly twice the no-CR value at later times, indicating that CRs enhance cold gas accretion and thereby feed central star formation.}
\label{f:SFR}
\end{figure*}

\begin{figure}
\includegraphics[width=\linewidth]{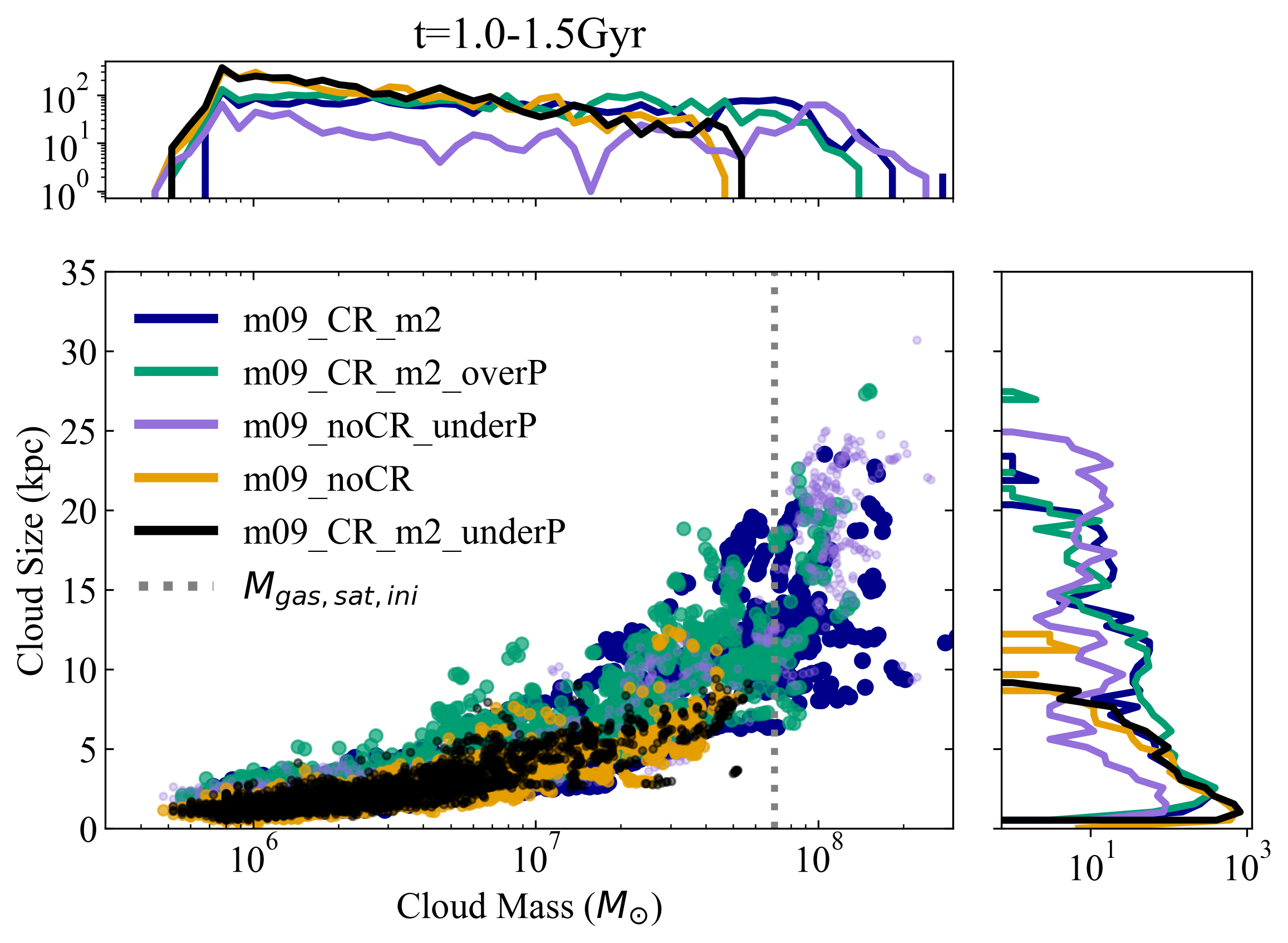}
\includegraphics[width=\linewidth]{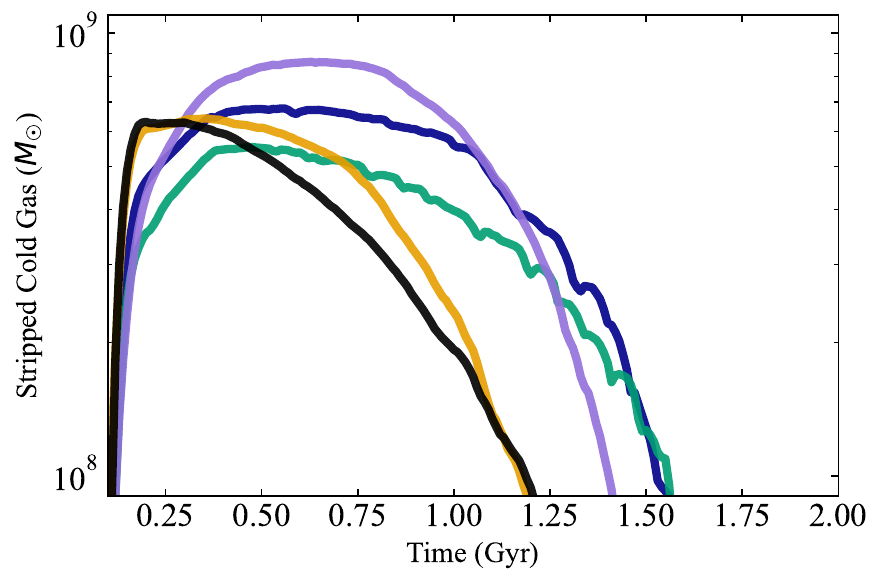}
\includegraphics[width=\linewidth]{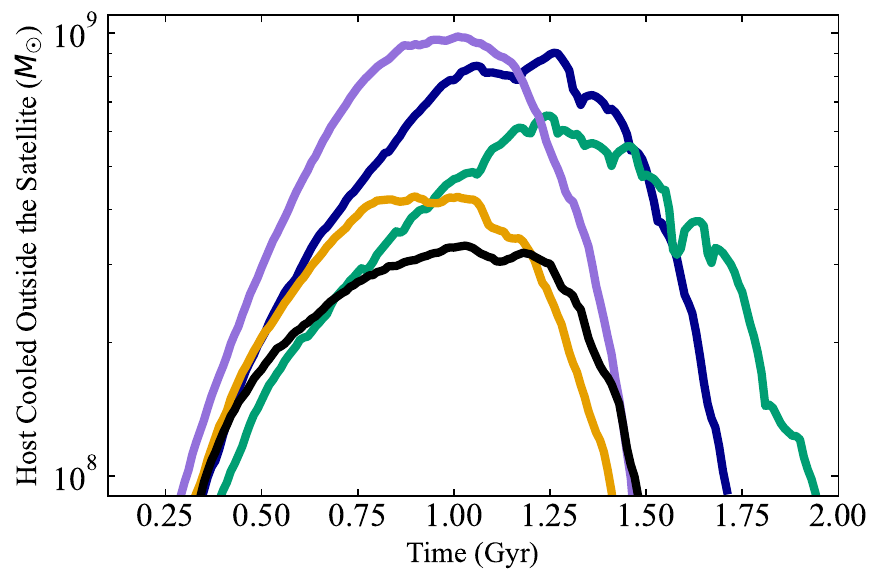}
\caption{Cloud growth and mixing-layer cooling tests isolating the effects of reduced thermal pressure from those of CRs. \textit{Top:} Cloud size and mass distributions for the fiducial CR run (\texttt{m09\_CR\_m2}; dark blue), overpressurized CR run (\texttt{m09\_CR\_m2\_overP}; green), underpressurized no-CR run (\texttt{m09\_noCR\_underP}; orange), and low-thermal-pressure CR run (\texttt{m09\_CR\_m2\_underP}; purple). \textit{Middle:} Time evolution of stripped gas; \textit{Bottom:} Time evolution of mixing-layer cooling. The \texttt{m09\_CR\_m2\_overP} case shows similar cloud growth to the fiducial CR run, indicating that CRs drive the enhancement independent of thermal pressure reduction. The \texttt{m09\_noCR\_underP} case shows only modest increases over the no-CR baseline, confirming that reduced thermal pressure alone cannot reproduce the CR-driven effect. Further lowering the thermal pressure in the CR run slightly increases cloud size and prolongs mixing-layer gas survival, but the dominant influence remains the presence of CRs.}
\label{f:temp_vs_CR}
\end{figure}

\subsection{Do CRs affect inflow rate?}
\label{S:inflow}
In this section, we assess whether the cold gas contributed by the satellite is capable of feeding the central galaxy. We restrict our analysis to the m09 run, as the stripping timescale for the m10 satellite exceeds 2 Gyr, and our high-CR simulation can not be analyzed beyond this point due to excessive halo cooling. We aim to answer two key questions: (1) What is the inflow rate of cold gas into the ISM? and (2) Can this cold gas supply the galaxy with sufficient material for star formation?

Figure~\ref{f:mass_flux} presents the time evolution of the net mass flux (where negative indicates inflow) of cold gas (defined as $T \leq 3 \times 10^4$\, K). The flux is computed across two spherical shells of 1 kpc thickness located at 80 kpc, as Satellites are in between 100-150 kpc (left panels) and 40 kpc as this is our boundary between CGM and inner galaxy (right panels) from the galactic center. The top and bottom panels represent flux contributions from satellite-origin and host-origin cold gas, respectively. 
We show instantaneous flux by light-shaded lines and time-integrated flux by dark-shaded lines.
We find that the instantaneous inflow rate increases with increasing CR pressure. For the time-integrated flux, the change of flux between low-CR and no-CR is minimal, while the mid-CR runs show higher flux than no-CR. In particular, the largest difference exists in the mass flux at the 40 kpc shell--the boundary between the CGM and the inner galaxy--where in the high-CR run, the peak instantaneous flux is enhanced by factors of $\sim2-4$ compared to the no-CR case for satellite and host gas, respectively. This suggests that CRs significantly facilitate the delivery of cold gas into the central galaxy.

Since the satellites are located between 100 and 150 kpc, cold gas begins to enter the 80 kpc shell earlier ($\sim$0.6 Gyr), while it reaches the 40 kpc shell slightly later ($\sim$0.8 Gyr), for both satellite and host gas. For satellite-stripped cold gas, the high-CR run shows a peak instantaneous inflow rate of 1.4 $M_\odot$yr$^{-1}$ at 80 kpc, which slightly increases to 1.75 $M_\odot$ yr$^{-1}$ at 40 kpc, indicating survival of these clouds. The host-derived cold gas flux significantly increases from 80 kpc to 40 kpc, suggesting that additional cooling in the mixing layers--amplified as the gas moves into denser regions-- enhances the cold gas mass. The time-integrated cold gas mass flux at 2 Gyr from satellites remains nearly constant at $\sim 6\times10^8\,\mathrm{M}_\odot$ between 80\,kpc and 40\,kpc, suggesting limited cloud destruction. In contrast, the integrated inflow mass flux from cooled host gas at 2 Gyr increases significantly, reaching from $\sim10^9\,\mathrm{M}_\odot$  to $\sim3\times10^9\,\mathrm{M}_\odot$ over the same radial range, indicating substantial cloud growth driven by cooling of the host gas. This behavior underscores the role of CRs in promoting both the survival and growth of cold gas clouds seeded by satellites and mixed with the ambient CGM.

However, previous studies have shown that CR can suppress gas inflow by providing additional pressure support. For instance, \cite{Butsky2020} demonstrated that when the CR pressure is comparable to the thermal pressure ($P_{\rm CR}/P_{\rm g} \sim 1$), the mass accretion rate can be reduced by nearly an order of magnitude. 
This suppression is not due to a decrease in the cold gas mass fraction (cloud survival or destruction), but rather a reduction in the inflow velocity due to increased CR pressure support against gravity, particularly in regions where the cooling time is long and CR pressure dominates.
Similarly, \citet{Su2020} showed that even moderate CR pressure is capable of suppressing cooling flows due to the extra pressure support it provides against gravity. \cite{Trapp2022} also found slightly lower mass fluxes in the CR simulations than in the no-CR case. In contrast, our results indicate that CRs can enhance the inflow of cold gas into the central galaxy, particularly in the high-CR case, where we observe inflow rates up to 2--8 times higher than those in the no-CR run at 40\,kpc. This enhancement arises from the increased survival of ram-pressure stripped gas and growth of the cold gas (not due to the change in the inward velocity) by additional cooling of the host halo gas within mixing layers, suggesting a different mode of CR-regulated accretion in satellite-rich environments.

Given the high inflow rates of cold gas, we next examine whether this gas can effectively contribute to star formation in the host galaxy. To assess this, we analyze the star formation rate (SFR) evolution of the host galaxy shown in the right panel of Figure~\ref{f:SFR}. Light-shaded curves show the raw SFR time series (we only show them for
no-CR, low-CR and high-CR in order to avoid a crowded plot), while bold curves indicate smoothed trends (moving average). We find that the SFR in the CR runs exhibits greater temporal fluctuations compared to the no-CR case. All the CR runs show an early suppression of SFR up to 1.2 Gyr due to extra-provided CR feedback, consistent with previous studies \citep{chan:2018.cosmicray.fire.gammaray,Su2020,hopkins:cr.transport.physics.constraints}. While the average SFRs in the CR runs are comparable (high CR runs; h and m2) or lower (low CR runs; m1, l) to the no-CR run in the early times, the instantaneous SFR in the high-CR simulation sometimes stands out with a SFR that is $2$-$3$ times higher than the others at some snapshots due to large fluctuations. However, in high-CR runs, at a later time ($\sim1.2$ Gyr when the inflow rate becomes high for high-CR), the burstiness and suppression in SFR decline more than in low-CR runs due to a sustained high influx of cold gas. The average SFR for high CR runs also increases to twice that of the no-CR run. We also plot the total cold gas mass flux at $40$ kpc in the left panel of Figure~\ref{f:SFR}. 
The high CR-run shows total net cold gas fluxes about a factor of 4 higher than the no-CR run, which aligns with the ratio of the SFRs at later times.
Therefore, this suggests that the large, coherent cold clouds supported by CR pressure not only survive infall but also deliver fuel to the galaxy, enabling enhanced star formation.  

In summary, CRs significantly enhance the inflow of cold gas into the central galaxy, with the high-CR run exhibiting inflow rates up to 2-8 times higher than the no-CR case at 40 kpc. This increased flux results from both improved survival of satellite-stripped gas and additional cooling-driven growth of host gas in mixing layers. Due to the agreement between the mass flux and central SFRs in Figure~\ref{f:mass_flux}, we argue that as this cold gas reaches the inner galaxy, it can fuel star formation by boosting instantaneous rates at least 2-5 times at some snapshot, along with an increase in the average SFR at a later time for the high-CR run, even though CRs suppress the earlier SFR.


\section{Discussion}
\label{S:Discussion}

\subsection{What is causing the clouds to grow: CGM temperature or CRs?}
In this paper, we have found that CRs impact cloud survival and the multiphase structure of the CGM, and we have argued that this is due to the direct influence of CR pressure on cloud properties.  However, the cooling of the CGM and subsequent growth of cold clouds could also be due to a cooler CGM allowing for a lower temperature of mixing-layer gas that can more rapidly cool. To assess whether the observed cloud growth is primarily driven by CRs or by the associated reduction in the thermal pressure and temperature of the CGM, we use the \texttt{m09\_CR\_m2} simulation as our fiducial case. In this run, CR pressure constitutes one-quarter of the total pressure, thereby reducing the thermal pressure to three-quarters of its no-CR value. A plausible alternative explanation for the enhanced cloud growth, in this case, is that the lower thermal pressure (and hence lower CGM temperature, due to the identical density profiles for our different runs) promotes more efficient cooling in the mixing layer, leading to cloud growth.  

To test this, we perform three controlled variations of the fiducial run (described in the final three rows in Table \ref{t:runs}). In the first, we increase the thermal pressure back to its original no-CR value while retaining the CR component, resulting in an overpressurized (by $25\%$) CGM (\texttt{m09\_CR\_m2\_overP}, compared to hydrostatic equilibrium; shown in green). If the enhanced cloud growth were solely due to the reduction in thermal pressure, no enhancement of growth should occur in this case. However, as shown in the top panel of Figure~\ref{f:temp_vs_CR}, the cloud growth (green) remains qualitatively similar to that in the fiducial \texttt{m09\_CR\_m2} run (dark blue). In the middle panel, the stripping rate in \texttt{m09\_CR\_m2\_overP} is somewhat higher than in \texttt{m09\_CR\_m2}, shifting the timescale for mixing layer cooling (shown in the bottom panel), yet the total amount of mixing layer cooling remains comparable between the two. This indicates that CRs directly contribute to cloud growth and survival.  

In the second variation, we lower the thermal pressure to three-quarters of its no-CR value but exclude CRs, producing an underpressurized (by $25\%$) halo (\texttt{m09\_noCR\_underP}; compared to hydrostatic equilibrium; shown in orange). If reduced thermal pressure alone were responsible for enhanced cloud growth, we would expect results similar to the fiducial CR run. Instead, as shown in the top panel of Figure~\ref{f:temp_vs_CR}, the resulting cloud population closely resembles that of the original no-CR run, with only a marginal increase in the size and mass of some clouds. This is consistent with the bottom panel, where a slightly greater mixing layer cooling is observed relative to the no-CR case, but still far less than in the fiducial CR run \texttt{m09\_CR\_m2}.  

Finally, we test the effect of further reducing the thermal pressure in the fiducial CR run from three-quarters to one-half of the no-CR value, yielding a total pressure equal to the under-pressurized no-CR run (\texttt{m09\_CR\_m2\_underP}; shown in purple). This further reduction only slightly increases the typical cloud size and mass and modestly prolongs the survival of mixing layer gas compared to the fiducial CR run.  

Overall, these tests demonstrate that although reduced thermal pressure can contribute to modest cloud growth and extended survival, the dominant mechanism driving the substantial enhancement seen in our simulations is the direct influence of CRs.

\subsection{Effect of CR transport: variation in diffusion coefficient and magnetic field}
\label{S:transport}
In this section, we will investigate whether CR transport has some effect on cloud growth. The CR diffusion and advection timescales are given by:

\[
t_{\mathrm{diff}} = \frac{R^2}{D} \approx 30 \, \mathrm{Gyr} \left( \frac{R}{100 \, \mathrm{kpc}} \right)^2 \left( \frac{D}{10^{29} \, \mathrm{cm}^2/\mathrm{s}} \right)^{-1}
\]

\[
t_{\mathrm{adv}} = \frac{L}{v} \approx 0.98 \, \mathrm{Gyr} \left( \frac{L}{100 \, \mathrm{kpc}} \right) \left( \frac{v}{100 \, \mathrm{km/s}} \right)^{-1}
\]

These scalings illustrate that diffusion is a significantly slower process compared to advection unless the diffusion coefficient is extremely large. For CRs injected at the center of the galaxy, diffusion alone is insufficient to transport them to distances of $\sim 100 \, \mathrm{kpc}$ within a Hubble time, unless $D \gg 10^{29} \, \mathrm{cm}^2/\mathrm{s}$. Advection, on the other hand, can reach such distances on Gyr timescales for typical outflow velocities. However, this timescale is still comparable to or longer than the timescale over which satellites experience significant ram-pressure stripping according to our simulations. 
Consequently, in simulations with very low initial CR pressure, the contribution from centrally injected CRs does not arrive in time to meaningfully affect the satellite environment. This explains the minimal CR impact observed in such low-CR scenarios.  While CRs are also injected by star-forming satellites, we find that their contribution is limited due to their low star formation rates. Thus, even local CR injection from satellites due to star formation alone is unlikely to dominate the dynamics, unless injected in the Black Hole and/or the bow shock, which we plan to explore in future work. 
Therefore, the effect of CR energy density is dominant over the CR feedback from the host and satellite. 

 Our simulations initialized with higher CR pressure, however, show a pronounced impact on satellite gas stripping, consistent with CRs providing additional pressure support and modifying mixing and cooling in the cloud-wind interface. Given the uncertainties in CR transport, we further explore this regime by running our high-CR run with reduced diffusion and another with lower $\beta$ (i.e., a stronger magnetic field). Since streaming proceeds at the Alfven speed, a higher $\beta$ corresponds to more efficient CR transport. Overall, we find that stronger CR transport (via diffusion and streaming) leads to a flattening of the CR radial profile both in the inner CGM of the host and in the vicinity of satellites. This redistribution enhances the CR fraction in the inner CGM ($\sim 60$ kpc) and around the satellites. 
 
\begin{figure*}
\includegraphics[width=0.49\linewidth]{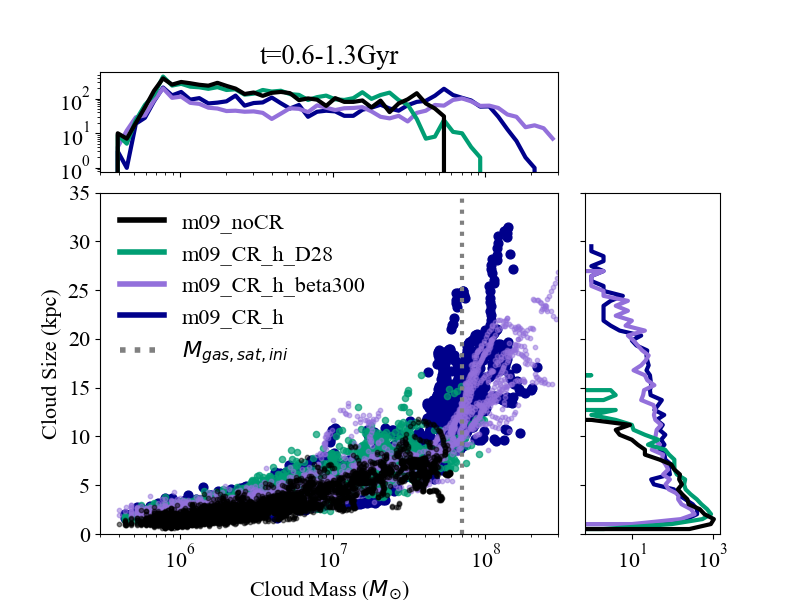}
\includegraphics[width=0.49\linewidth]{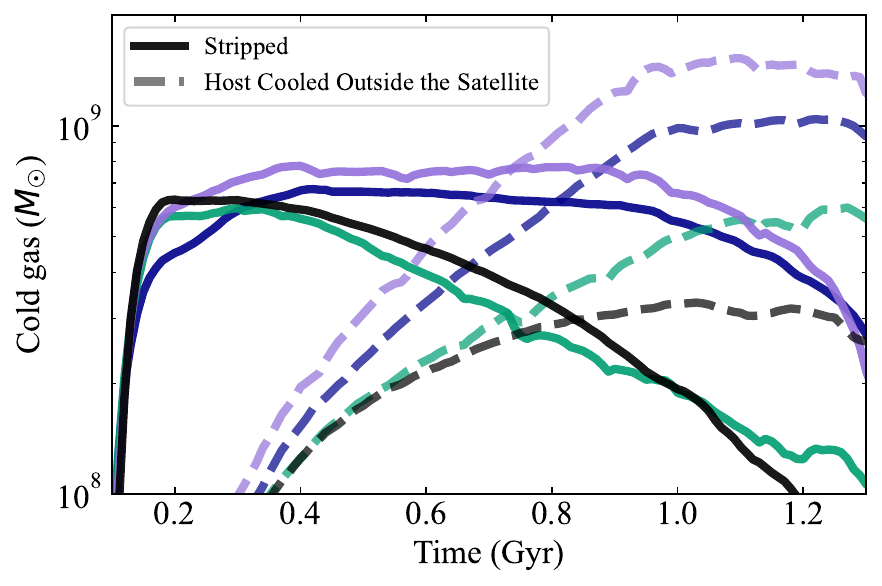}
\caption{Illustration of the role of CR transport in regulating cloud survival and mixing in the halo. We show the variation in diffusion coefficient and magnetic field with our fiducial high-CR run, with left panel showing cloud size-mass distribution at later time and right panel showing the stripped and mixing layer cold gas time evolution. A high CR  alone is insufficient; efficient CR transport—via a high diffusion coefficient and/or strong magnetic field—is also required for CRs to become dynamically important. The effect is more sensitive to the diffusion coefficient than to the magnetic field strength.}
\label{f:transport}
\end{figure*}

We show that the case with a lower diffusion coefficient of $\sim10^{28}$ cm$^2$ s$^{-1}$ results in more rapid satellite stripping (right panel of Figure \ref{f:transport}) as more CRs are trapped within the satellite, leading to higher CR pressure inside the satellite, which will act in concert with ram pressure. Also, the lower $D$ results in the CRs spreading less efficiently into the clouds and their surroundings and consequently reduces the impact of CRs. Moreover, it also reduces the diffusion of CRs from small radii to the host CGM, resulting in a lower CR pressure within $r \lesssim 60$ kpc. As a result, the \texttt{m09\_CR\_h\_D28} run exhibits less mixing layer cooling than our fiducial run with a higher diffusion coefficient, as shown in the right panel of Figure~\ref{f:transport}. This trend is also evident in the left panel, where the cloud size and mass distributions indicate that, in the low-$D$ case, CRs do not significantly inflate the clouds relative to the no-CR run, leading to only modest size increases. Nevertheless, the enhanced cooling of the mixing layer still drives some mass growth compared to the no-CR case.

From an observational perspective, 
\citet{chan:2018.cosmicray.fire.gammaray} 
showed that $D \sim 3\times 10^{29} \, \mathrm{cm}^2 \, \mathrm{s}^{-1}$, or slightly lower values, are required to match the observed $\gamma$-ray luminosities of $L_\ast$ systems. Similarly, \citet{Roy2022} found that the $\gamma$-ray luminosity observed in both the disk and extended ($\sim 100$ kpc) halo of M31 can only be explained if $D \gtrsim 10^{29} \, \mathrm{cm}^2 \, \mathrm{s}^{-1}$. Consequently, $D \sim 10^{28} \, \mathrm{cm}^2 \, \mathrm{s}^{-1}$ is lower than what has been inferred in ISM by \cite{chan:2018.cosmicray.fire.gammaray}, given the current $\gamma$-ray constraints. We also do not explore diffusion coefficients substantially higher than $3 \times 10^{29} \, \mathrm{cm}^2 \, \mathrm{s}^{-1}$, since such values would imply unphysically rapid CR transport across the halo, again inconsistent with the $\gamma$-ray constraints discussed by \citet{chan:2018.cosmicray.fire.gammaray}. Although a diffusion coefficient of $10^{28} \, \mathrm{cm}^2 \, \mathrm{s}^{-1}$ is not physically motivated by current $\gamma$-ray observations, we highlight the role of CR transport by changing $D$ by an order of magnitude from our fiducial value.


The impact of CR transport is further illustrated in our high-magnetic-field run, where we adopt $\beta = 300$ (the ratio of thermal to magnetic pressure) while keeping the fiducial diffusion coefficient fixed. As shown in the right panel of Figure~\ref{f:transport}, this run exhibits a stripping rate comparable to the low-$B$ fiducial case, but with more mixing layer cooling (the difference in the dashed lines). The left panel shows that the cloud size distribution remains similar to that of the low-$B$ runs, although the mass distribution extends to larger values in the high-$B$ case. This suggests that, unlike earlier simulations \citep[e.g.][]{Thomas} without CRs, where magnetic draping suppresses the mixing, the inclusion of CRs might introduce a competing effect that could change the outcome.

Therefore, we find that stronger CR transport giving rise to a higher CR pressure (around the satellites and in the host CGM within $r \lesssim 60,\mathrm{kpc}$) leads to stronger mixing-layer cooling and longer survival of ram-pressure–stripped cold gas, as also demonstrated in the main text where we explicitly vary the CR fraction. These results again underscore the importance of CR transport in shaping cloud evolution.  Therefore, we emphasize that not only do we need a high CR fraction in the halo, but we also need efficient CR transport (high diffusion coefficient and/or high magnetic field) for CRs to become dynamically important in regulating cloud survival and mixing. However, our results indicate that this dependence is more sensitive to the diffusion coefficient than to the magnetic field strength. We will investigate the variation of CR transport on the CGM in more detail in our future work.  
\subsection{Limitations of this work}
We note several limitations of our setup. First, our simulations are idealized and do not include a cosmological context; in particular, they lack cosmological accretion and large-scale evolution, both of which could drive additional turbulence and supply cold gas to the outer halo. Second, our satellite configuration is simplified and does not reflect the actual distribution of Milky Way satellites, although the satellite mass range we adopt is broadly consistent with that observed in the Milky Way. Finally, CR transport in the CGM remains highly uncertain due to limited observational constraints, and the poorly known CGM magnetic field further compounds these uncertainties, as CR transport and energy losses depend sensitively on its properties. We address the uncertainty in CR transport in Section \ref{S:transport}.

\section{Conclusion}
\label{S:Conclusion}

In this work, we explore the role of cosmic rays (CRs) in shaping the cold phase of the circumgalactic medium (CGM), focusing specifically on ram-pressure stripped gas from satellite galaxies in a Milky Way-like environment in an idealized simulation. Our key findings are summarized below:

\begin{itemize}
\item \textbf{CRs alter cloud morphology:} CR pressure makes ram-pressure stripped cold gas clouds more extended and diffuse (puffier), increasing their surface area and suppressing fragmentation (Figure \ref{f:snapshots}, \ref{f:cloud}).

\item \textbf{Enhanced mixing-layer cooling:} The larger surface area of CR-inflated clouds facilitates more efficient mixing with the surrounding hot CGM, leading to increased cooling and cloud mass growth (Figure \ref{f:cloud} and \ref{f:time_evo}).

\item \textbf{Higher cold gas budget:} The total cold gas mass in the CGM is up to 4$\times$ higher in the high-CR runs compared to no-CR runs, especially around the satellite stripping epoch (Figure \ref{f:time_evo}).

\item \textbf{Greater covering fractions:} CRs lead to larger, more coherent clouds that increase the sky covering fraction of cold gas, with high-CR runs showing up to 8$\times$ higher covering fractions in external galaxy projections (Figure \ref{f:CF}).

\item \textbf{Increased cold gas inflow:} The growth and survival of cold clouds in CR-rich environments result in significantly higher inflow rates into the central galaxy - up to 2--8 times greater than in the no-CR case at 40\,kpc (Figure \ref{f:mass_flux}).

\item \textbf{Boosted star formation:} The enhanced cold gas supply in CR runs drives a $\sim$2$\times$ increase in the instantaneous and average star formation rate at later time, particularly in the high-CR scenarios (Figure \ref{f:mass_flux}, \ref{f:SFR}).

\end{itemize}
In conclusion, our study demonstrates that CR-dominated ram-pressure stripped clouds can evolve into significantly larger and more massive structures. Not only do we find that CRs in the CGM will enhance the cold gas content of the host halo, but they will affect the central galaxy as well by increasing the cold gas inflow rate and therefore the star formation rate.  Although we have focused on clouds stripped from satellites, we expect the effect of CRs on cold gas survival to be true for clouds formed within or injected into the CGM by any other mechanism as well.  Finally, our work highlights the importance of including physically accurate CR models in simulations of galaxy formation and evolution. 

\newpage
\section*{Acknowledgments}
We thank Greg Bryan, Phil Hopkins, Peng Oh, Claude-André Faucher-Giguére, and the FIRE collaboration for useful discussions and suggestions. MR acknowledges support from the CCAPP fellowship at The Ohio State University and ACCESS allocations of PHY240003. MR acknowledges the Aspen Center for Physics and Simons Foundation, as part of this work was performed there, which is supported by a grant from the Simons Foundation (1161654, Troyer). KS acknowledges support from the Black Hole Initiative at Harvard University, which is funded by grants from the John Templeton Foundation and the Gordon and Betty Moore Foundation, and acknowledges ACCESS allocations TG-PHY220027 and  TG-PHY220047 and Frontera allocation AST22010.  The computations in this work were run at facilities supported by the Bridges, University of Pittsburgh and Anvil, Purdue University.
\appendix
\section{Convergence Test}
\begin{figure*}
\includegraphics[width=0.49\linewidth]{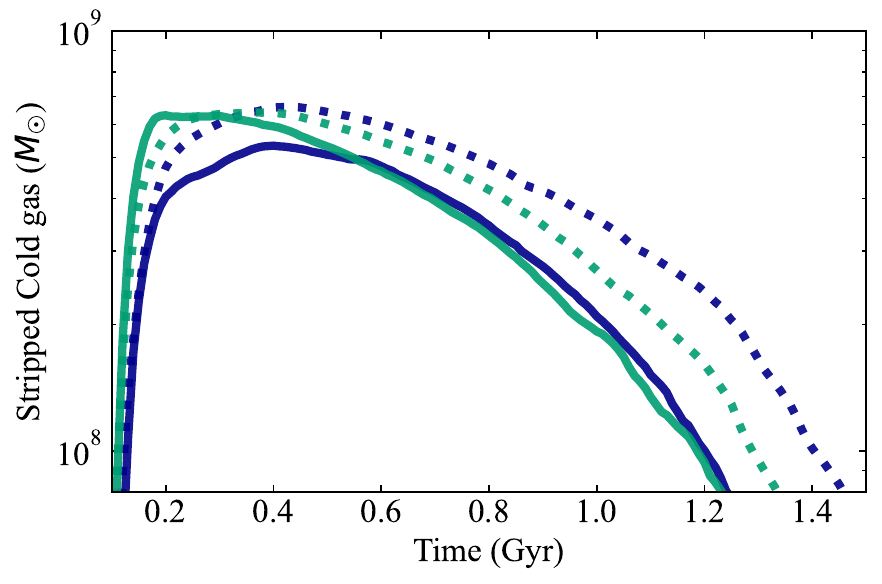}
\includegraphics[width=0.49\linewidth]{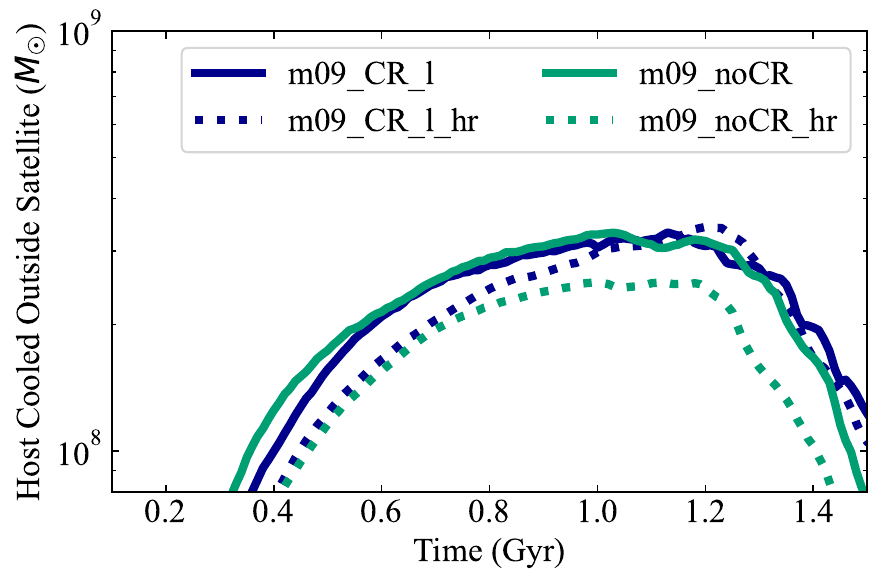}
\caption{Left: Mass of stripped cold gas from the satellite as a function of time for different resolution runs. Right: Mass of host gas that cools outside the satellite in the same runs. Higher resolution enhances the stripping rate,  but the qualitative trends and conclusions remain unchanged.}
\label{f:res}
\end{figure*}

We have also performed higher-resolution (hr) runs, shown by the dotted lines in Figure \ref{f:res}. We expect that the hr runs will better resolve denser, colder gas compared to the low-resolution (lr) runs. 
Upon initial inspection, we find that the cold gas mass in the hr runs is qualitatively similar to that in the lr runs and exhibits the same trends with satellite mass, indicating that our general results are robust against resolution changes.
In the hr runs, cold, dense gas is initially more difficult to strip, resulting in more stripped gas for a longer time compared to the lr runs, for both CR and no-CR cases (left panel). The mixing layer cooling is also reduced in the hr runs relative to the lr runs at early times, as the hr runs resolve smaller structures that are more easily disrupted compared to larger clouds with longer survival times \citep{Gronke2018,Fielding2022}. However, at later times, the CR runs show similar levels of mixing layer cooling for both hr and lr runs, suggesting convergence for CR runs. In contrast, the no-CR hr run continues to exhibit less mixing layer cooling even at late times. This implies that increasing resolution, even in the presence of very little CRs, enhances mixing layer cooling relative to the no-CR case, further tightening our conclusion.


\vspace{0.3cm}
\section*{Data Availability statement}
The data supporting the plots within this article are available on reasonable request to the corresponding author. A public version of the GIZMO code is available at \href{http://www.tapir.caltech.edu/~phopkins/Site/GIZMO.html}{\textit{http://www.tapir.caltech.edu/$\sim$phopkins/Site/GIZMO.html}}.

\normalsize

\label{lastpage}
\bibliography{main}
\bibliographystyle{aasjournal}
\end{CJK}
\end{document}